\documentclass[apj]{emulateapj}
\submitted{Accepted by AJ, May 30, 2005}
\usepackage{lscape}
\newcommand{\Bp}{\ensuremath{B\!_p}}
\newcommand{\sUB}{\ensuremath{\sigma_{U\!-\!B}}}

\newcommand\ubvri{\mbox{$U\!BV\!R_\mathrm{C}I_\mathrm{C}$}}
\newcommand\ugriz{\ensuremath{ugriz}}
\newcommand\ri{\mbox{$R\!-\!I$}}
\begin{document}
\title{The SDSS View of the Palomar-Green Bright Quasar Survey}
\author{
Sebastian Jester,\altaffilmark{1} \email{jester@fnal.gov}
Donald P. Schneider,\altaffilmark{2} \email{dps@astro.psu.edu}
Gordon T. Richards,\altaffilmark{3} \email{gtr@astro.princeton.edu}
Richard F. Green,\altaffilmark{4} \email{green@noao.edu}
Maarten Schmidt,\altaffilmark{5} \email{mxs@astro.caltech.edu}
Patrick B. Hall,\altaffilmark{6} \email{phall@yorku.ca}
Michael A. Strauss,\altaffilmark{3} \email{strauss@astro.princeton.edu}
Daniel E. Vanden Berk,\altaffilmark{2} \email{danvb@astro.psu.edu}
Chris Stoughton,\altaffilmark{1} \email{stoughto@fnal.gov}
James E. Gunn,\altaffilmark{3} \email{jeg@astro.princeton.edu}
Jon Brinkmann,\altaffilmark{7} \email{jb@apo.nmsu.edu}
Stephen M. Kent,\altaffilmark{1,8} \email{skent@fnal.gov}
J. Allyn Smith,\altaffilmark{9,10} \email{jasman@nis.lanl.gov}
Douglas L. Tucker,\altaffilmark{1} \email{dtucker@fnal.gov}
and Brian Yanny\altaffilmark{1} \email{yanny@fnal.gov}
}

\altaffiltext{1}{Fermi National Accelerator Laboratory, M.S. 127,
  P.O. Box 500, Batavia, IL 60510}
\altaffiltext{2}{Department of Astronomy and Astrophysics, The Pennsylvania State University, 525 Davey Laboratory, University Park, PA 16802.}
\altaffiltext{3}{Princeton University Observatory, Peyton Hall, Princeton, NJ 08544.}
\altaffiltext{4}{Kitt Peak National Observatory, National Optical Astronomy Observatories, P.O. Box 26732, 950 North Cherry Avenue, Tucson, AZ 85726}
\altaffiltext{5}{Department of Astronomy, MC 105-24, California Institute of Technology, 1200 East California Boulevard, Pasadena, CA 91125.}
\altaffiltext{6}{Department of Physics and Astronomy, York University, 4700 Keele St., Toronto, Ontario M3J 1P3, Canada}
\altaffiltext{7}{Apache Point Observatory, P.O. Box 59, Sunspot, NM 88349.}
\altaffiltext{8}{Department of Astronomy and Astrophysics, The University of Chicago, 5640 South Ellis Avenue, Chicago, IL 60637.}
\altaffiltext{9}{Department of Physics and Astronomy, University of Wyoming, P.O. Box 3905, Laramie, WY 82071.}
\altaffiltext{10}{Los Alamos National Laboratory, P.O. Box 1663, Los Alamos, NM 87545.}


\begin{abstract}
We investigate the extent to which the Palomar-Green (PG) Bright
Quasar Survey (BQS) is complete and representative of the general
quasar population by comparing with imaging and spectroscopy from the
Sloan Digital Sky Survey. A comparison of SDSS and PG photometry of
both stars and quasars reveals the need to apply a color and magnitude
recalibration to the PG data.  Using the SDSS photometric catalog, we
define the PG's parent sample of objects that are not main-sequence
stars and simulate the selection of objects from this parent sample
using the PG photometric criteria and errors.  This simulation shows
that the effective $\ub$ cut in the PG survey is $\ub<-0.71$, implying
a color-related incompleteness.  As the color distribution of bright
quasars peaks near $\ub=-0.7$ and the 2-$\sigma$ error in $\ub$ is
comparable to the full width of the color distribution of quasars, the
color incompleteness of the BQS is approximately 50\% and essentially
random with respect to $\ub$ color for $z<0.5$.  There is, however, a
bias against bright quasars at $0.5 < z < 1$, which is induced by the
color-redshift relation of quasars (although quasars at $z>0.5$ are
inherently rare in bright surveys in any case).  We find no evidence
for any other systematic incompleteness when comparing the
distributions in color, redshift, and FIRST radio properties of the
BQS and a BQS-like subsample of the SDSS quasar sample.  However, the
application of a bright magnitude limit biases the BQS toward the
inclusion of objects which are blue in $g-i$, in particular compared
to the full range of $g-i$ colors found among the $i$-band limited
SDSS quasars, and even at $i$-band magnitudes comparable to those of
the BQS objects.
\end{abstract}


\defcitealias{BQS}{SG83}
\defcitealias{PG}{GSL86}

\section{Introduction}\label{s:intro}

\subsection{The Palomar-Green Bright Quasar Survey}
\label{s:intro.BQS}

The Bright Quasar Survey \citep[BQS;][hereafter referred to as
\citetalias{BQS}]{BQS} is the set of quasars detected in the
Palomar-Green (PG) survey of ultraviolet excess objects
\citep[][hereafter referred to as \citetalias{PG}]{PG}.  The PG
survey selects UV excess objects with $\ub<-0.46$ (corresponding to
$\ub < -0.44$ for quasars with the mean color difference between stars
and quasars assumed by \citetalias{BQS}) brighter than an effective
limiting magnitude of $B_\mathrm{lim}=16.16$.  The BQS was the first
large-area homogeneous quasar survey, and it remains the largest-area
survey for bright quasars at $B<16$.  Because of the UV excess
criterion, it is primarily sensitive to quasars at redshifts up to
$z=2.2$, where the Lyman-$\alpha$ line enters the $B$-band and gives
quasars very red $\ub$ colors ($\ub > 0$).

\citetalias{BQS} inferred strong luminosity evolution of quasars from
a combination of the BQS and fainter small-area quasar surveys.
However, a number of authors find a substantially larger number of
bright UV excess quasars, implying that the BQS is substantially
incomplete.  For example, \citet{GMlFea92} reported that the bright
quasar surface density from the Edinburgh Quasar Survey is three times
that found by the BQS, leading to very different volume densities of
low-redshift high-luminosity quasars and hence different results on
the evolution of the quasar luminosity function \citep{GM98}.  By
contrast, \citet{WCBea00} find that the BQS has a completeness of 68\%
compared to the Hamburg-ESO quasar survey (HEQS).  Next to the BQS,
the HEQS is the largest-area survey reaching similarly bright
magnitudes as the BQS, but only overlaps part of the PG survey area as
it surveyed the Southern hemisphere.

Some authors have raised concerns that the BQS incompleteness may be
systematic with respect to optical color or radio properties.  For
example, \citet{WP85} have suggested that the paucity of BQS quasars
in the redshift interval $0.5 < z < 1.0$ may be caused by a BQS bias
against redder quasars (or ``quasars with yellow \ub\ colors'', as
designated by \citealt{WP85}) --- in this redshift interval, the
passage of the MgII line through the $B$ filter results in a $\ub$
color that is 0.2 mag redder than at lower and higher redshifts.
However, \citet{LFEea97} have argued that a cut of $\ub<-0.44$ is red
enough to avoid this bias.  Based on independent photometry of the
luminous ($M_B<-24$) BQS objects, \citet{WP85} also suggested that
some of the PG limiting magnitudes (which varied from plate to plate)
might have been ``much fainter than stated''.  

There are also concerns about a systematic incompleteness with respect
to radio properties.  \citet{MRS93} found that 50\% of $z<0.5$ BQS
quasars are steep-spectrum radio-loud objects; they cautioned that
this number might be spuriously high if the BQS selection favored the
inclusion of radio-luminous objects.  This cautionary note was
interpreted by \citet{GKMea99} and others as \emph{suggesting} a
radio-dependent incompleteness, casting doubts on the results obtained
from statistical analyses of the radio properties of BQS objects.

In fact, PG quasars are often considered to be the archetype optically
selected quasar sample, in particular when studying the properties of
optically selected quasars at other wavelengths.  Many other optically
selected quasar surveys are now available which reach much fainter
magnitudes and higher redshifts, use more general selection criteria,
and generally have a much larger sample size (e.g., the Palomar
Transit Grism Survey by \citealp{SSG94} with 90 objects at $2.75 < z
< 4.75$ over an area of 61.5\,deg$^2$; the Large Bright Quasar Survey
by \citealp{HFC95}, 1055 objects with $16 < B_J < 18.9$ at $0.2 < z <
3.4$ over 454\,deg$^2$; COMBO-17 by \citealp{WWBea03}, 192 objects
with $17<R<24$ at $1.2 < z < 4.8$ over 0.78\,deg$^2$; the 2dF QSO
redshift survey by \citealp{CSBea04}, 23 338 objects with $16 <
b_\mathrm{J} < 20.85$ at $z<3$ over 674\,deg$^2$); in particular, the
SDSS quasar survey \citep{qso_ts} has cataloged nearly 50,000 quasars
so far, including of order 500 at $z>4$ \citep{SFHea03,SHGea05} and,
in a separate search, 16 at $z>5.7$ \citep{FHRea04}.  Each of these
surveys will be more or less biased towards or against objects with a
particular class of SEDs.

Because of the historical importance of the BQS in general, and its
key role in anchoring the bright end of the local quasar luminosity
function in particular, we investigate here whether the BQS suffers
from any systematic incompleteness.  By ``incompleteness'', we refer
both to lack of objects that in fact pass the survey's magnitude and
color limits, but also to the extent to which the survey is
representative of quasars satisfying the broadest definition of the
term (objects showing non-stellar continua and broad emission lines).
We do so by comparing PG and Sloan Digital Sky Survey
\citep[SDSS;][]{Yorea00} photometry of PG sources
(\S\ref{s:SDSSphot}), by considering the completeness of the PG UV
excess sample relative to UV excess sources from SDSS
(\S\ref{s:simPG}), and by comparing properties of BQS objects to those
of ``BQS-like'' quasars from the SDSS (\S\ref{s:compPGSDSS}), with
special attention to radio properties in \S\ref{s:comp.radio.bias}.
We also consider which part of the quasar population found by the SDSS
are selected by the BQS criteria.  In our discussion section, we
compare our findings to previous investigations of the BQS
incompleteness (\S\ref{s:disc.compBQS}) and consider the biases
induced by the selection criteria of the BQS and other quasar surveys.
We summarize our findings in \S\ref{s:sum}.

\subsection{Selection of objects in the BQS}
\label{s:intro.selBQS}

\begin{figure}
\plotone{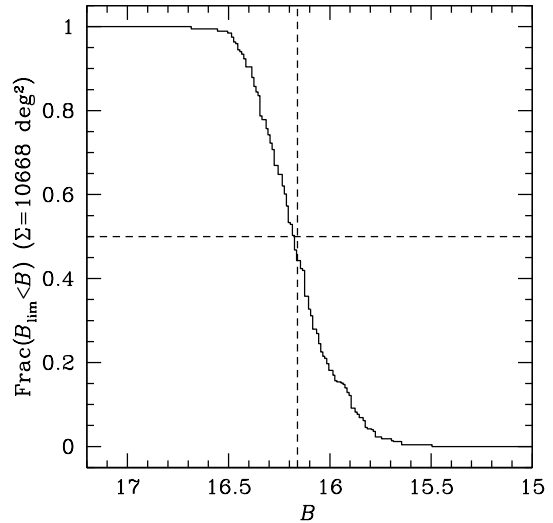}
\caption{\label{f:intro.BQS.Blim}Area-weighted cumulative histogram of
limiting magnitudes in the complete PG sample \citepalias[using the
areas and limiting magnitudes of PG survey plates from][Table 1]{PG}.
The effective limiting magnitude $B_\mathrm{eff}=16.16$ given by
\citetalias{BQS} and \citetalias{PG} is very close to the
area-weighted median limiting magnitude as indicated by the dashed
lines.}
\end{figure}

For reference, we briefly review the construction of the BQS sample
from \citetalias{BQS} and \citetalias{PG}.  The PG survey selects UV
excess objects on double exposures of baked IIa-O photographic plates
through filters $U$ (using Schott UG-2) and $B$ (GG-13) obtained using
the Palomar 18\,inch (46\,cm) Schmidt telescope. The exposure times
were adjusted to obtain equal image densities for objects with
$\ub=-0.46$ (with the mean color difference between the spectral
energy distribution [SED] of stars and quasars assumed by
\citetalias{BQS}, this translates to $\ub = -0.44$ for quasars).  The
magnitudes were calibrated using photoelectric calibration exposures
of local standards. The calibration was stabilized by fitting a model
of star counts as function of galactic coordinates.  The limiting $B$
magnitude varied from plate to plate; Figure~\ref{f:intro.BQS.Blim}
shows the distribution of limiting magnitudes.  \citetalias{BQS} give
an effective limiting magnitude of $B<16.16$ and photometric errors of
$\sigma_B=0.27$ and $\sUB=0.24$. These were later revised by
\citetalias{PG} to $\sigma_B=0.34$ and $\sUB=0.39$.  The area covered
by the complete PG survey was 10,668\,deg$^2$.

UV excess candidates passing the photometric selection criteria were
observed spectroscopically to determine quasar redshifts and to remove
main-sequence objects (i.e., all objects determined to have \ion{Ca}{2} K
lines) that had been scattered into the sample.  Conversely, the
spectroscopic observations were started before the photometric
calibration had been finalized, therefore some objects that had been
confirmed spectroscopically as objects that are not main-sequence
stars (hereafter called ``off-main sequence objects'') were retained
in the PG sample even if the final photometric calibration formally
caused them to fail the photometric selection criteria.

The BQS consists of 114 PG objects, which were required to have
``dominant starlike appearance on blue prints of the 48 inch (1.2 m)
Schmidt Sky Atlas'' (the Palomar Sky Survey) and ``broad emission
lines with substantial redshift'' \citepalias{BQS} \citep[we use the
  most up-to-date version of the BQS object list from][see notes to
  Tab.~\ref{t:BQSinSDSS}]{KSSea94}.  Like all bright quasar surveys,
the redshift distribution of BQS objects peaks at low redshift, with a
median $z$ of 0.176, and 88 out of 114 objects (77\%) at $z<0.5$.

\subsection{The SDSS}
\label{s:intro.sdss}

The Sloan Digital Sky Survey \citep{Yorea00,edr} is a photometric and
spectroscopic survey of the extragalactic sky at high galactic
latitude visible from the northern hemisphere.  In this paper, we use
the data set published as Data Release 3
\citep[DR3;][]{DR3}\footnote{Available on-line at
  \url{http://www.sdss.org/dr3/}}.  The DR3 photometric catalog covers
5282\,deg$^2$, while the DR3 spectroscopic catalog covers
4188\,deg$^2$. The SDSS uses a wide-field camera \citep{Gunea98} to
obtain $\ugriz$ photometry
\citep{ugriz,LGS99,Hogea01,LIKea01,ugrizstandards,QA} with a
  photometric accuracy of 2--3\%, a comparable precision of placing
  the photometry on the AB magnitude scale \citep{OG83,dr2}, and with
  an astrometric accuracy of better than 0\farcs1 \citep{astrom}.  The
  SDSS is also a spectroscopic survey of complete samples of galaxies
  \citep{lrg_ts,gal_ts} and quasars.  Quasar candidates are selected
  from SDSS photometry using an algorithm \citep{qso_ts} targeting
  objects with non-stellar colors or that are point-source optical
  counterparts of FIRST \citep{BWR95} radio sources.  A tiling
  algorithm ensures uniformity and efficiency of the allocation of
  fibers to quasar candidates and other spectroscopic targets
  \citep{tiling}.  A manually vetted quasar catalog based on DR3 data
  akin to that based on DR1 \citep{SFHea03} is given by
  \citet{SHGea05} and contains over 46,000 quasars.

In this paper, we use $\Omega_{\mathrm{m}} = 0.3, \Omega_\Lambda =
0.7$, and $H_0 = 70$\,km/s/Mpc.

\section{The photometric properties of PG sources in the SDSS}
\label{s:SDSSphot}
To evaluate the photometric completeness of the BQS, we perform
separate comparisons of the PG and SDSS photometry for stars and
quasars\footnote{We use the term ``quasar'' here, but it should be understood
as being synonymous with ``QSO'' without any implication about
luminosity or presence of radio emission}.  The stellar photometry
will be used to assess the photometric accuracy and precision of the
PG survey using SDSS observations of PG objects.  The quasar
photometry transformations will be used to define a sample of SDSS
quasars satisfying the BQS color selection criteria.

The SDSS photometric system is based on a set of standard stars which
includes stars as blue as PG objects \citep{ugrizstandards}.  The PG
survey reports photometry transformed to the Johnson $U$ and $B$
system.  Since a subset of the SDSS standard stars is taken from the
Landolt standard fields with Johnson-Kron-Cousins \ubvri\ photometry
\citep{Lan73,Lan83,Lan92}, it is straightforward to establish color
and magnitude transformations between the SDSS and the Landolt system.
However, these transformations are not appropriate for quasars because
of their different spectral shapes.  Therefore, we determine separate
color and magnitude transformations for quasars by performing
synthetic photometry of composite quasar spectra in the SDSS and
Landolt systems. We describe each process in turn.

\subsection{Photometric transformation for stars}
\label{s:phot.startrans}

\citet{ugriz} and \citet{ugrizstandards} give color and magnitude
transformations (synthetic and observed) between the Landolt system
and the system of the United States Naval Observatory (USNO) 1.0\,m
telescope (designated $u^\prime g^\prime r^\prime i^\prime
z^\prime$). We transform the USNO magnitudes of the SDSS standard
stars to the system of the SDSS~2.5\,m telescope (i.e., the SDSS
$ugriz$ system) using the equations given by \citet{JEGphoteq}. We use
only those standard stars with $U-B < 0$ and perform linear least
squares fits to determine the transformations from the SDSS system to
$U$ and $B$. The resulting coefficients are given in
Table~\ref{t:phototrans}.  Ideally, separate transformations should be
derived for stars of different spectral classes because of differences
in the strength of the Balmer lines, in particular between white
dwarfs and the remaining blue stars. Indeed, synthetic photometry
suggests that DA white dwarfs have a $B-g$ that is greater by about
0.1 than that of the remaining white dwarfs and hot subdwarfs;
however, we do not have enough stars with both accurate spectral
classifications and Johnson photometry to confirm this using observed
magnitudes (this is true both for the SDSS standards and for the
subset of PG stars with photoelectric photometry).  In any case, the
standard stars that have been used include the full range of colors
observed in the PG survey.

For reference in work with objects redder than $U-B=0$, we also give
the coefficients obtained using all SDSS standard stars with $\ri <
1.15$ (stars with $\ri < 1.15$ have different color transformations
than those with $\ri \ge 1.15$, but there is no Landolt photometry of
a sufficient number of SDSS standards with $\ri \ge 1.15$ for us to
derive reliable transformations), as well as coefficients for
transformations between other filters.

\begin{deluxetable*}{lrcll}
\tablewidth{0pt}
\tablecaption{\label{t:phototrans}Transformations between \ugriz\ and
  \ubvri}
\tablehead{\colhead{Sample} & \multicolumn{3}{c}{Magnitude/Color
    Transformation} & \colhead{RMS residuals}}
\startdata
\multicolumn{5}{c}{\ugriz\ to \ubvri}\\
\tableline
Quasars at $z\leq 2.1$ (synthetic) & $\ub$ & = & $0.75 (u-g) -0.81$ & 0.03\\
&  $\bv$ & = & $0.62 (g-r) + 0.15 $ & 0.07\\
&  $\vr$ & = & $0.38 (r-i) + 0.27 $ & 0.09\\
&  $\ri$ & = & $0.72 (r-i) + 0.27 $ & 0.06\\
&    $B$ & = & $g + 0.17 (u-g) + 0.11$ & 0.03\\
&    $V$ & = & $g - 0.52 (g-r) -0.03 $ & 0.05\\
\tableline
Stars with $\ri < 1.15$ and $\ub < 0$ & $\ub$ & = & $0.77 (u-g) -
0.88$ & 0.04\\
&  $\bv$ & = & $0.90 (g-r) + 0.21 $ & 0.03\\
&  $\vr$ & = & $0.96 (r-i) + 0.21 $ & 0.02\\
&  $\ri$ & = & $1.02 (r-i) + 0.21 $ & 0.01\\
&    $B$ & = & $g + 0.33 (g-r) + 0.20$ & 0.02\\
&    $V$ & = & $g - 0.58 (g-r) -0.01 $ & 0.02\\
\tableline
All stars with $\ri < 1.15$ & $\ub$ & = & $0.78 (u-g) - 0.88$ & 0.05\\
&  $\bv$ & = & $0.98 (g-r) + 0.22 $ & 0.04\\
&  $\vr$ & = & $1.09 (r-i) + 0.22 $ & 0.03\\
&  $\ri$ & = & $1.00 (r-i) + 0.21 $ & 0.01\\
&  $B$ & = & $g + 0.39 (g-r) + 0.21 $ & 0.03\\
&  $V$ & = & $g - 0.59 (g-r) -0.01 $ & 0.01\\
\tableline
\multicolumn{5}{c}{\ubvri\ to \ugriz}\\
\tableline
Quasars at $z\leq 2.1$ (synthetic) & $u-g$ & = & $1.25 (\ub) +
1.02$ & 0.03\\
&  $g-r$ & = & $0.93 (\bv) - 0.06 $ & 0.09\\
&  $r-i$ & = & $0.90 (\ri) - 0.20 $ & 0.07\\
&  $r-z$ & = & $1.20 (\ri) - 0.20 $ & 0.18\\
&    $g$ & = & $V + 0.74 (\bv) -0.07$ & 0.02\\
&    $r$ & = & $V - 0.19 (\bv) -0.02 $ & 0.08\\
\tableline
Stars with $\ri < 1.15$ and $\ub < 0$ & $u-g$ & = & $1.28 (\ub) +
1.14$ & 0.05\\
&  $g-r$ & = & $1.09 (\bv) - 0.23 $ & 0.04\\
&  $r-i$ & = & $0.98 (\ri) - 0.22 $ & 0.01\\
&  $r-z$ & = & $1.69 (\ri) - 0.42 $ & 0.03\\
&    $g$ & = & $V + 0.64 (\bv) -0.13$ & 0.01\\
&    $r$ & = & $V - 0.46 (\bv) +0.11 $ & 0.03\\
\tableline
All stars with $\ri < 1.15$ & $u-g$ & = & $1.28 (\ub) + 1.13$ & 0.06\\
&  $g-r$ & = & $1.02 (\bv) - 0.22 $ & 0.04\\
&  $r-i$ & = & $0.91 (\ri) - 0.20 $ & 0.03\\
&  $r-z$ & = & $1.72 (\ri) - 0.41 $ & 0.03\\
&    $g$ & = & $V + 0.60 (\bv) -0.12$ & 0.02\\
&    $r$ & = & $V - 0.42 (\bv) +0.11 $ & 0.03\\
\enddata
\tablecomments{Transformations for quasars are derived from synthetic
photometry of an updated version of the quasar composite from
\citet{Berea01} using DR1 data as well as the red and reddened quasar
composites from \citet{Ricea03}. Transformations for stars are derived
from $u^\prime g^\prime r^\prime i^\prime z^\prime$ photometry (on the
system of the USNO~1.0\,m telescope) of Landolt standards given by
\citet{ugrizstandards} after transformation to $ugriz$ (the system of
the SDSS 2.5\,m survey telescope) using the equations given by
\citet{JEGphoteq}.  \citet{ugrizstandards} showed that stars with $\ri
< 1.15$ have different color transformations than those with $\ri \ge
1.15$; as there is no Landolt photometry of a sufficient number of
SDSS standards with $\ri \ge 1.15$, we restrict ourselves to $\ri <
1.15$ here.}
\end{deluxetable*}

\subsection{Photometric transformation for quasars}
\label{s:phot.qsotrans}

\begin{figure*}
\plottwo{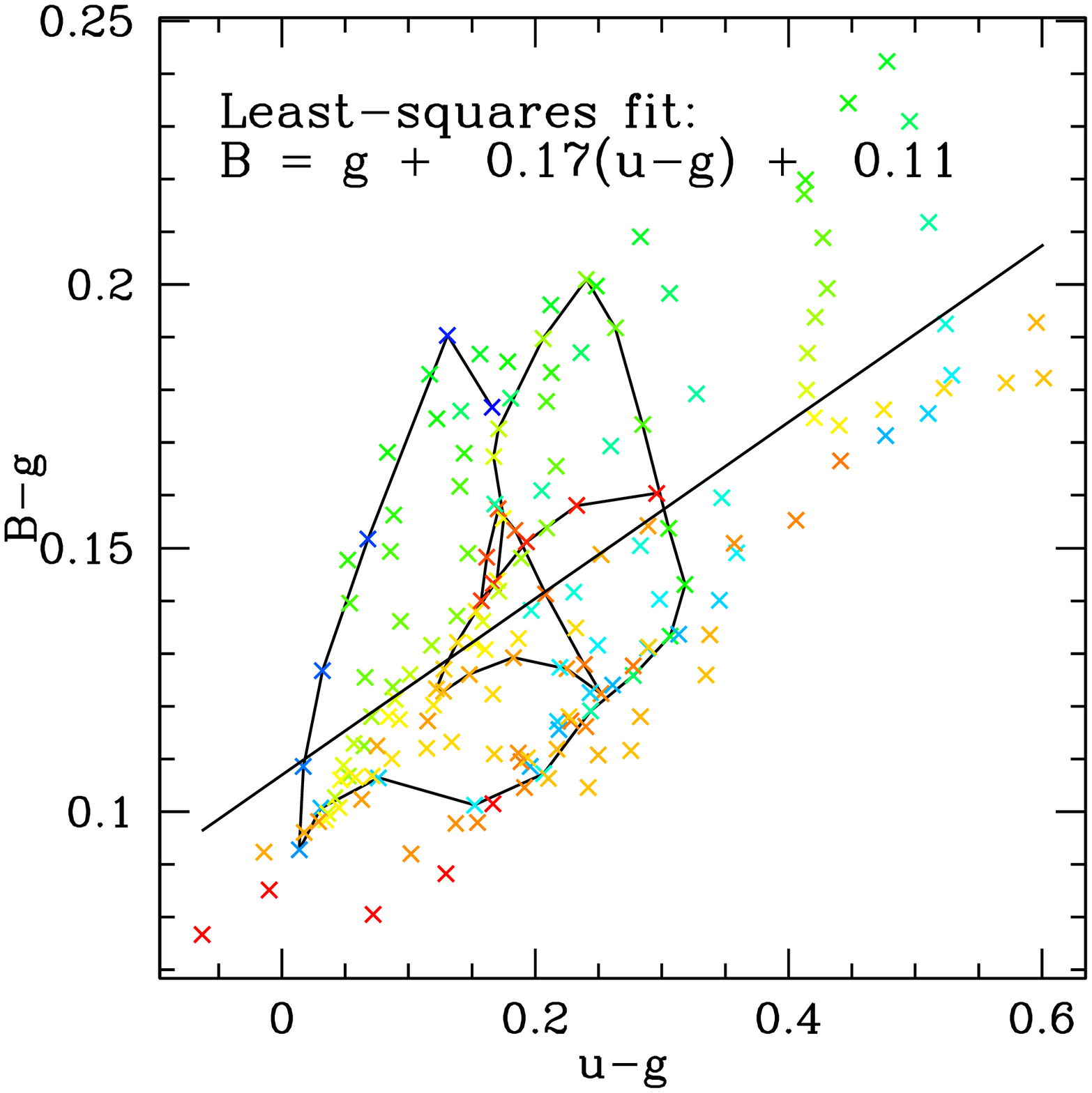}{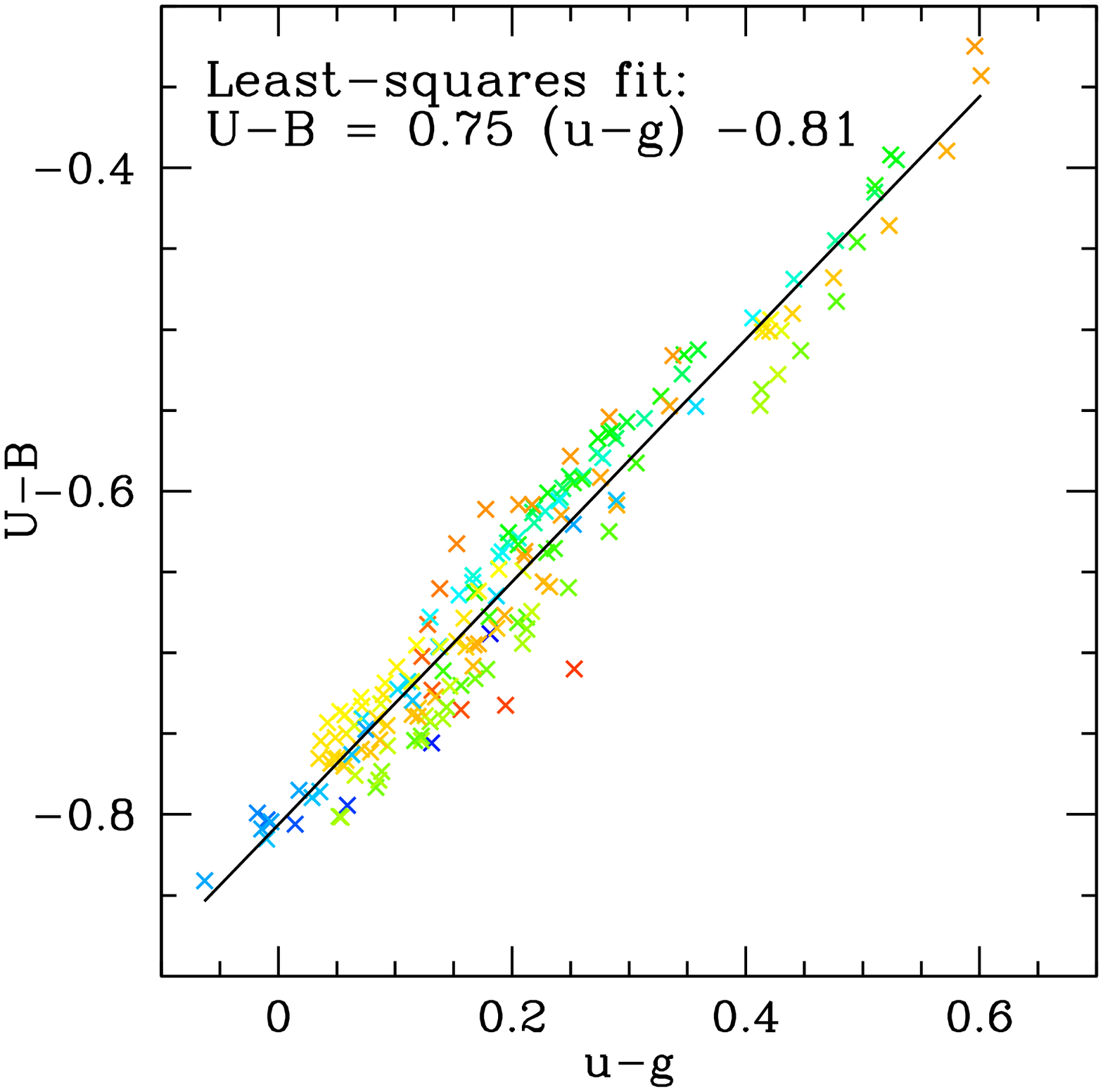}
\caption{\label{f:qsotrans}Transformation of quasar colors and
  magnitudes from SDSS $ug$ to Landolt $U\!B$ obtained from synthetic
  photometry of quasar composites at redshifts from 0 to 2.1 in steps
  of 0.05.  Points are color-coded by redshift in the on-line
  edition. The points belonging to the updated version of the
  composite from \citet{Berea01} are connected by a solid line in the
  left-hand plot to illustrate the change in color and magnitude
  difference as a quasar spectrum is redshifted; the remaining points
  are for the red and reddened composites from \citet{Ricea03}. The
  diagonal solid lines show the best-fit straight line.  The RMS
  scatter about the best fit is less than 0.03 in both cases.}
\end{figure*}

To account for the different spectral shapes of stars and quasars, in
particular the presence of strong emission lines, we derive separate
transformation equations for quasars.  Since all quasars are variable
at optical wavelengths, it would be necessary to use contemporaneous
SDSS and $U\!B$ photometry of a sample of quasars to derive adequate
transformations from observations.  As such observations are not
available, we perform synthetic photometry of an updated version of
the composite from \citet{Berea01} using DR1 data (\citealp{dr1};
changes in spectrophotometric calibration of SDSS data introduced in
DR2 do not change the transformations appreciably) and the composite
spectra from \citet{Ricea03} for quasars with different intrinsic
colors and reddening.  We use the STSDAS Synphot package under the
PyRAF scripting environment\footnote{Obtained from
\url{http://www.stsci.edu/resources/software\_hardware/}} and the
Landolt filter curves provided with that package as well as the SDSS
transmission curves\footnote{Available at
\url{http://www.sdss.org/dr2/instruments/imager/\#filters}} to
determine color and magnitude differences for each composite. We
compute a set of colors each at redshifts ranging from 0 to 2.1 in
steps of 0.05 for the \citet{Berea01} composite and 0.3 to 1.6 for the
remaining composite spectra, which cover a smaller range of rest-frame
wavelengths.  For the comparison with BQS photometry, we fit \ub\ and
$B-g$ as a function of both $u-g$ and $g-r$ for all composites
simultaneously.  As both \ub\ and $B-g$ are more tightly correlated
with $u-g$ than with $g-r$, we use the fits as a function of $u-g$ to
derive the transformations.  The synthetic photometry results and
best-fit lines are shown in Figure~\ref{f:qsotrans}.  The resulting
coefficients are also given in Table~\ref{t:phototrans}, which again
also contains transformations for the remaining SDSS and $\ubvri$
filters for reference. 

The RMS scatter in the transformed quantities is 0.03, smaller than
the plot suggests to the eye, although the $B-g$ residuals can be as
large as 0.1 and show some systematic behavior with redshift.  These
systematics are caused by the presence of emission lines; the
transformations as given have a precision of 0.06 or better in
transforming pure power-law spectra.  The \ub\ transformation is only
slightly different from that for blue stars, but the $B-g$
transformation differs significantly --- not only do we fit $B-g$ as a
function of $u-g$ instead of $g-r$, but even if we had used $g-r$, the
transformations would differ by between 0.06 to 0.2 magnitudes in the
$g-r$ range of quasars ($-0.1$ to 0.5). With the transformations in
hand, we now consider the SDSS photometry of PG sources.

\subsection{SDSS observations of PG sources}
\label{s:sdss.match}

We are interested in the SDSS photometry of PG sources, both to
determine whether there are any systematics in the photometric
calibration of the PG survey, and to compare the PG quasars to those
selected by the SDSS quasar survey with its much wider color selection
criteria and magnitude cut in a much longer-wavelength band.  To this
end, we cross-match the PG catalog with the SDSS DR3 photometric
catalog.  The DR3 covers an area of 5282 square degrees; out of this
area, roughly 3300 square degrees are contained within the PG area of
10,668 square degrees.

Because of the relatively large astrometric uncertainty of the PG
survey \citepalias[roughly 9\arcsec\ RMS in every coordinate with
  systematic offsets of less than 2\arcsec; see \S III in][]{PG} and
the different observing epochs, the closest SDSS object by position is
not necessarily the correct match.  We therefore use a generous
positional matching radius of 2 arcminutes and employ cuts on the $B$
magnitude and \ub\ color transformed from SDSS photometry, accepting
only objects within 1 magnitude of the PG photometry and with $\ub <
-0.2$.  This procedure results in up to 3 SDSS matches per PG object;
where multiple matches are present, the bluest object in \ub\ among
them is always the correct match, the other matched objects being
redder stars.  Every PG object has a counterpart detected in the
SDSS. In our analysis below, we do not use photometry of matches which
are saturated in the SDSS or have otherwise unreliable photometry
using the flag checking recommendations for point sources from the
SDSS web
site\footnote{\url{http://www.sdss.org/dr3/products/catalogs/flags.html}};
this affected 9 objects.

We find 52 matches for BQS quasars in the DR3 photometric catalog
which are shown in Table~\ref{t:BQSinSDSS}. All of these are either
identified as quasar candidates by the SDSS target selection, have a
cosmetic defect that excludes them from target selection (this
affected 5 objects; the criteria for cosmetic defects are described in
detail in \citealp{qso_ts}; \citealp{SOE} estimate that the fraction
of quasars missed due to image defects and blends is approximately
4\%), or are brighter than the SDSS spectroscopic bright limit of
$i>15$ which has been introduced to avoid fiber cross-talk and
saturation in the spectrograph. Thus, every BQS quasar is either
targeted by the SDSS, or we understand why it was not targeted.
However, even if target \emph{selection} is complete, spectra of
individual PG quasars may still be missing in the SDSS quasar survey
if an object was not targeted by one of the pre-final versions of the
target selection algorithm, or because of less reliable photometry in
the TARGET version of photometry \citep[see discussion of TARGET and
  BEST photometry in][\S3]{dr2}.  This affects one PG object for which
DR3 photometry is available: the final quasar target selection
algorithm correctly identified PG~1012+00 as a quasar candidate, but
the TARGET version of the photometry does not recognize it as a quasar,
merely as a counterpart to a ROSAT source.  The quasar resides in a host
galaxy with complex morphology (perhaps a merging system), with a
bright galactic nucleus approximately 2\arcsec\ away from the quasar.
The galactic nucleus was targeted as a galaxy and obtained a fiber
(Plate/MJD/Fiberid 270/51909/586), while the quasar itself did not,
because galaxy targets obtain a fiber with higher priority than ROSAT
matches \citep[see][]{edr,tiling}.

The photometric calibration is most easily assessed by considering
only non-variable stars, which are also much more numerous than the PG
quasars.  We obtain 466 SDSS matches of PG stars \citepalias[i.e.,
  objects classified as white dwarf or hot subdwarf by][]{PG} with
clean SDSS photometry in the DR3 area.  The SDSS-derived
color-magnitude diagram of these stars shows that the White Dwarfs and
hot subdwarfs of types sdO and sdB in the PG survey cover similar
ranges in color and magnitude.

\subsubsection{SDSS observations of PG stars: the accuracy of the PG
  photographic magnitudes}
\label{s:phot.sdss.accuracy}

Using the transformations established in \S\ref{s:phot.startrans}, we
derive the $U\!B$ photometry of non-variable PG stars from their
matches in SDSS data.  \citetalias{PG} report a $B$-band magnitude
(transformed to the standard Johnson system, which we assume to be
identical to the Landolt system here) derived from the photographic
photometry for every object in the survey, as well as a photoelectric
$B$ magnitude (which we will designate \Bp) and \ub\ color for a small
subset of the stars. The photographic \ub\ colors were not reported by
\citetalias{PG} because they were deemed less reliable as indicator of
each source's spectral type than the spectroscopic information.
However, one of the authors (RG) retrieved archival notes with the
photometric \ub\ colors for most of the BQS objects (i.e., the quasars
from the PG survey), we will compare those to SDSS photometry in
\S\ref{s:phot.sdss.colors} below.

\citetalias{PG} give an error of $\sigma_{\Bp}=0.05$ for the
photoelectric magnitudes, comparable to the accuracy of the SDSS CCD
photometry ($\sigma_u=0.03,\sigma_g=0.02$).  Both of these measures
are much more accurate than the photographic-plate derived magnitudes
with their quoted error of $\sigma_B=0.29$ \citepalias{PG}.

\begin{figure*}
\epsscale{0.75}
\plotone{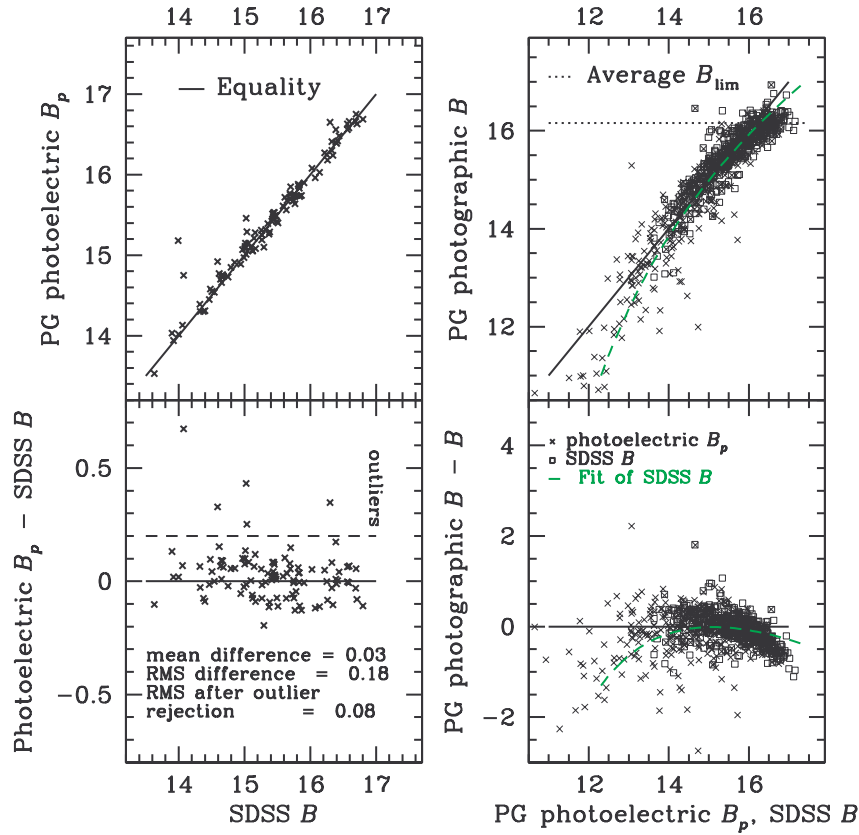}
\epsscale{1}
\caption{\label{f:phot.PGstars_Bfit}Comparison of SDSS and PG
  photometry for non-variable stars.  \emph{Left}, comparison of SDSS
  $B$ to PG photoelectric \Bp\ of 104 stars. \emph{Right}, crosses
  show PG photoelectric \Bp, squares show SDSS $B$ against PG $B$
  photometry derived from photographic plates.  In both cases, the
  upper panel shows the direct comparison of the magnitudes with the
  solid diagonal line indicating equality, while the lower panel shows
  the difference as a function of the more accurate magnitude (points
  above the dashed line in the lower left-hand panel have been
  rejected as outliers in the RMS determination). The dashed green
  lines in the right-hand side panels show the best-fit cubic
  (Equation~\ref{eq:Bfit}) describing the SDSS-derived or PG
  photoelectric $B$ magnitudes as function of the PG photographic $B$
  magnitude.  This fit adequately describes the calibration difference
  between PG photographic photometry and the Landolt system, while the
  more intuitive fit of the PG photographic magnitude as function of
  the PG photoelectric and SDSS $B$ magnitudes (not shown) is biased
  towards fainter photographic $B$ magnitudes by the removal of
  objects with photographic magnitudes fainter than the PG plate
  limits (the horizontal dotted line in the upper right-hand panel
  shows the effective limiting magnitude of 16.16).}
\end{figure*}

We begin with a comparison of the PG and SDSS magnitudes of these
stars in Figure~\ref{f:phot.PGstars_Bfit} (left-hand panels).  We
first compare the PG photoelectric \Bp\ and the SDSS $B$ magnitudes of
the 104 stars in the overlap sample which also have photoelectric PG
photometry.  The mean difference of $-0.03$ is comparable to the
photometric accuracy of the SDSS data.  The observed RMS difference of
0.18 is much larger than the $\sigma_{\Bp}=0.05$ given by
\citetalias{PG}.  However, after iterative rejection of $3\sigma$
outliers, the RMS difference drops to 0.075, as expected when
considering the difference of two quantities where each has an error
of 3\%-5\%.  The different RMS is thus caused by a non-Gaussian error
distribution in the PG photoelectric photometry.  A similar comparison
of photoelectric and SDSS \ub\ shows a mean and median offset of 0.01
and a surprisingly small RMS difference of 6\% (even without outlier
rejection).

We next compare the photographic-plate derived $B$ magnitudes to
SDSS-derived B magnitudes for all non-variable PG stars matched in the
SDSS, as well as to the PG photoelectric \Bp, where available
(right-hand panels in Figure~\ref{f:phot.PGstars_Bfit}). The
comparison shows that there are systematic differences between the
``low-accuracy'' PG photographic magnitudes and both of the
``high-accuracy'' magnitudes, the SDSS $B$ and the photoelectric \Bp\
magnitudes.  The PG photographic magnitudes are too bright by about
0.2 magnitudes on average at $B\approx 17$, and similarly too bright
at $B\approx 13$.  At $B\approx15$, the photographic magnitudes agree
with those from the SDSS, with a smooth transition from either
extreme, although there is a much larger scatter at the bright end.
The original photographic calibration approximated the photographic
S-curve by a linear fit in the magnitude range $14 < B < 16$,
consistent with the small calibration differences we find within this
magnitude range and the larger differences outside it.  There are no
discernible systematic trends of this calibration difference with the
\ub\ color. This calibration difference changes not only the
magnitudes of the PG objects (and hence the limiting magnitudes), but
also the inferred scatter between the PG and other photometry. We
therefore wish to recalibrate the PG $B$ magnitudes using SDSS and/or
PG \Bp\ photometry.

This recalibration \emph{cannot} be done simply by fitting the PG
photographic $B$ magnitudes as function of the high-accuracy SDSS $B$
and PG \Bp\ magnitudes because the machine used to digitize the
photographic plates did not report measurements of any objects fainter
than the plate limit (which varied from plate to plate).  Therefore,
the points shown in the right two panels in
Figure~\ref{f:phot.PGstars_Bfit} are restricted to lie below the line
$B=B_\mathrm{lim}$ for each PG plate (the dotted line shows the
average $B_\mathrm{lim}$ of 16.16).  Hence, no PG data are available
for those objects which have been scattered out of the PG sample, and
the fit is biased towards artificially bright photographic $B$ values
at the faint end.

To allow a correction of the limiting magnitudes of each PG field, we
instead perform a fit of the inverse relation, i.e. the SDSS or \Bp\
magnitude as function of PG photographic magnitude. In this case, the
censoring is applied in the independent variable, resulting in a less
biased fit, provided that errors in both coordinates are taken into
account.  We obtain the following fit:
\begin{eqnarray}
B & = &  15.1364 + 0.9584\,(B_\mathrm{PG}-15) \nonumber \\
     &   & -0.1605\,(B_\mathrm{PG}-15)^2 -0.0160\,(B_\mathrm{PG}-15)^3
\label{eq:Bfit}
\end{eqnarray}
where $B_\mathrm{PG}$ designates the photographic PG magnitude. This
fit is shown as green dashed line in Figure~\ref{f:phot.PGstars_Bfit}.
This fit closely follows a non-parametric fit obtained by determining
the median photographic magnitude as function of the high-precision
magnitudes in bins of size $\Delta B= 0.1$.  

We expect this calibration difference to be responsible for some of
the scatter between photographic and other $B$ magnitudes, so that the
error of $\sigma_B = 0.29$ should be an overestimate of the actual
error.  However, the residuals of the corrected photographic $B$ do
not have a distribution that is appreciably narrower than the original
residuals; in fact, both the original and the corrected residuals have
an RMS of 0.34.  Nevertheless, use of the fit clearly reduces the
systematic calibration errors.

\subsubsection{SDSS observations of BQS quasars: the accuracy of the PG
  photographic colors and $B$-band variability}
\label{s:phot.sdss.colors}

\begin{figure}
\plotone{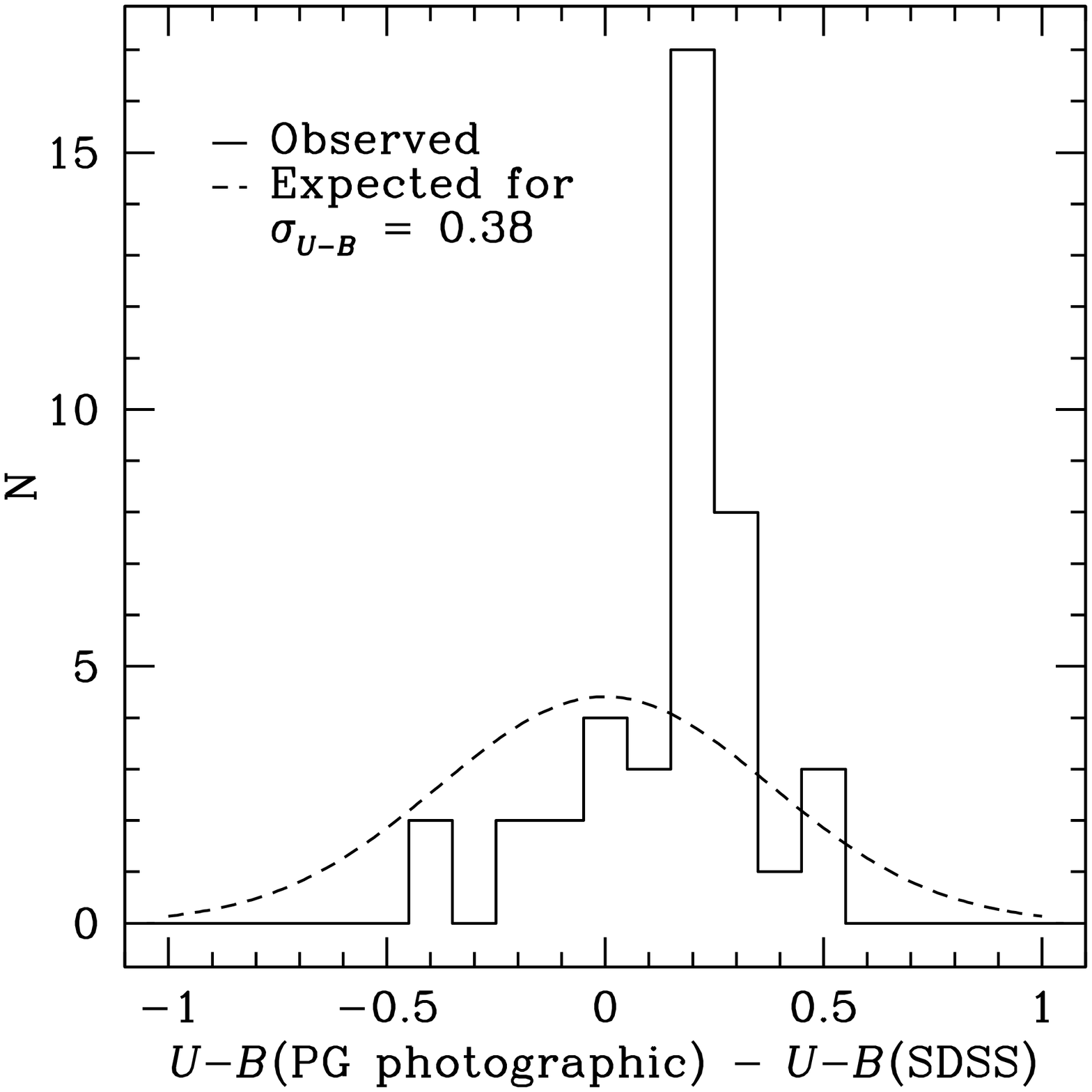}
\caption{\label{f:phot.UBdiff}Distribution of the difference between
\ub\ colors of BQS objects derived from SDSS and PG photometry.  The
difference histogram is sharply peaked at $\Delta\ub=0.2$, with a
median difference of 0.19 and an RMS of 0.21.  The dashed Gaussian
indicates the expected difference histogram for the \ub\ error of 0.38
quoted by \citetalias{PG}.}
\end{figure}

We now turn to the comparison of SDSS and PG \ub\ colors, where the
latter are available.  Figure \ref{f:phot.UBdiff} shows the
distribution of the difference between SDSS and PG-derived \ub\ for
the 47 BQS objects for which clean SDSS photometry is available. The
distribution of residuals is sharply peaked at $\Delta(\ub) =
(\ub)_\mathrm{PG}-(\ub)_\mathrm{SDSS} = 0.2$, with tails to
$\pm0.5$. The RMS \ub\ difference is 0.21 and is dominated by
photometric errors.  (Since the PG magnitude measurements are based on
double exposures of the same plate, the PG $U$ and $B$ measurements
are effectively contemporaneous, as are the SDSS observations in $u$
and $g$.)

\citetalias{PG} only performed photoelectric photometry of
non-variable stars, but not of any BQS objects; however, any color
variability over the epoch difference between PG and SDSS will be
negligible compared to the PG photographic color error
\citep{VBWKea04}, so that the color comparison is not affected by
variability.  Thus, judging from the peak in the $\ub$ difference
histogram, the PG-recorded photographic \ub\ colors are too red by
about 0.2 magnitudes. As a consequence, we expect that the true color
cutoff of the objects in the PG survey is in fact about 0.2 magnitudes
\emph{bluer} than assumed.

\begin{figure}
\plotone{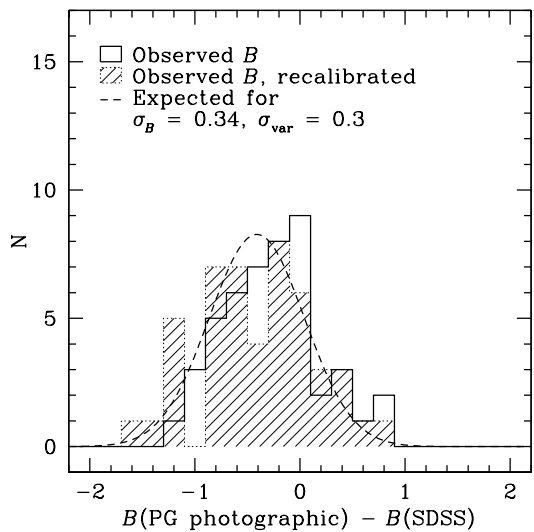}
\caption{\label{f:phot.Bdiff}Distribution of the difference between
$B$ magnitudes of BQS objects derived from SDSS and PG photometry.
The unshaded histogram shows the distribution of differences between
SDSS and PG photographic $B$ magnitudes as reported by \citetalias{BQS},
while the shaded histogram shows the distribution of differences after
correcting the PG photographic magnitudes using the fit derived above
(Equation~\ref{eq:Bfit}).  The recalibrated magnitudes have a mean
offset of $B(\mathrm{SDSS}) - B(\mathrm{PG}) = -0.40$, and a median
offset of $-0.32$.  The RMS difference is 0.54 magnitudes. Both offset
and RMS appear reasonably consistent with the offset anticipated for an
RMS quasar variability of 0.3 magnitudes over the 30-year epoch
difference between SDSS and PG, as indicated by the dashed Gaussian curve.}
\end{figure}

We compare SDSS and photographic $B$ magnitudes of the quasars in our
overlap sample in Figure~\ref{f:phot.Bdiff}.  On average, the quasars
are fainter by 0.4 magnitudes at the SDSS epoch, with an RMS
difference of 0.54 magnitudes. Given an RMS difference $\sigma$, the
expected magnitude offset is $\sigma^2$ times the logarithmic slope of
the number-magnitude counts, or about $2 \sigma^2$.  Thus, the
expected offset is about 0.5 magnitudes, consistent with the
variability amplitude of 0.3 magnitudes expected over the
approximately 30-year epoch difference between SDSS and PG
observations \citep{HSWea01}. The variability amplitude found here
also agrees roughly with the structure function for quasars presented
by \citet{dVBW03}.

In conclusion, we have found two key differences between the PG
photographic photometry on the one hand, on which the PG color and
magnitude selection is based, and SDSS and PG photoelectric photometry
on the other hand: there is an offset between the PG photographic and
the true $B$-band magnitudes which varies systematically with
magnitude, as well as an offset of 0.2 magnitudes between the
photographic and true \ub\ color (objects appear redder in the PG
survey than they are), for which no systematic variation with
magnitude could be established.  To double-check the validity of these
conclusions, we now simulate the PG survey's selection of objects from
its parent sample as observed with the SDSS.

\section{Completeness of the PG sample relative to UV excess sources
  from the SDSS}
\label{s:simPG}

We quantify the PG sample incompleteness by selecting SDSS object that
pass the PG photometric selection criteria.  Inclusion in the PG
catalog of UV excess objects furthermore required that objects pass a
spectroscopic confirmation as off-main sequence object or quasar. This
process led to the exclusion of 1125 main-sequence objects that had
been scattered into the photometric sample. SDSS spectroscopic target
selection explicitly rejects objects with the colors of white dwarfs
to improve the selection efficiency of quasar candidates
\citep{qso_ts}, so that a similar spectroscopic confirmation is not
available for all objects passing the PG criteria in SDSS photometry.
However, the photometric calibration of the SDSS is sufficiently
accurate to allow a clean separation of main-sequence and PG-like UV
excess objects in the $\ub$ against $B$ color-magnitude diagram.
Essentially all objects with $\ub < -0.3$ that are not saturated in
SDSS photometry are off the main sequence (the magnitude at which
objects are saturated of course depends on the seeing; the brightest
unsaturated object has a PSF magnitude $r=12.1$, and objects as faint
as $r=14$ can contain saturated pixels).

From this parent sample, we can directly simulate the expected color
and magnitude distribution of objects in the PG sample, in the
following manner.  Using the positions of the plate centers and
limiting magnitudes from Table~1 in \citetalias{PG}, we determine in
which PG survey plates each SDSS object is contained, and hence the PG
limiting magnitude for each object (for simplicity, we approximate the
shape of PG survey plates as perfect circles of area 59.14 square
degrees, i.e., with radius 4\fdg33876).  For objects covered by multiple
plates with different PG limiting magnitudes, we use the faintest one.
We transform SDSS magnitudes into the $U\!B$ system using the
transformations derived above (we apply the quasar transformations
from \S\ref{s:phot.qsotrans} to those objects with a quasar spectrum
in the SDSS, and the stellar transformations \S\ref{s:phot.startrans}
for all others).  For every SDSS object, we also determine whether it
has been included in the PG catalog.  In this way, we can compare the
completeness of the PG survey \emph{relative} to the SDSS; i.e., we
can determine whether there are any color or magnitude systematics to
the incompleteness.  We are not making any statements about the
\emph{absolute} surface density of UV excess objects in this section.

\begin{figure*}
\epsscale{0.75}
\plotone{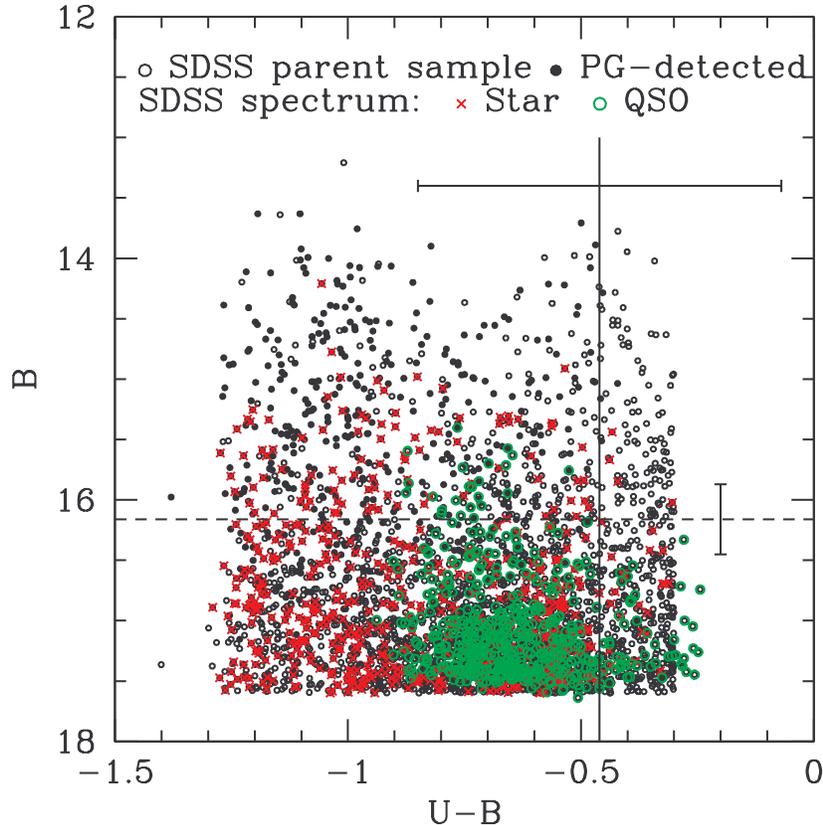}
\epsscale{1}
\caption{\label{f:simPG.UBB}Color-magnitude diagram showing the parent
  sample of the PG survey of UV excess objects selected from SDSS
  photometry. We show all SDSS objects with clean photometry,
  $\ub<-0.3$ and $B<17.6$ located within a PG survey plate (i.e., less
  than 4\fdg33876 from a plate center).  Filled circles represent
  objects with PG detections.  Colored symbols indicate SDSS
  spectroscopic classification where an SDSS spectrum is available:
  red crosses for stars (predominantly white dwarfs and hot subdwarfs)
  and green circles for quasars.  The SDSS spectroscopic survey has so
  far identified 546 quasars in this area.  The vertical solid line
  and the associated error bar indicate the nominal PG color cut of
  $-0.46$ ($-0.44$ for quasars) and the error on the $\ub$
  photographic color, while the dashed horizontal line and its error
  bar show the average limiting $B$ magnitude of 16.16 and the
  photographic $B$ magnitude error.  All objects were \emph{selected}
  from SDSS photometry assuming the photometric transformations for
  stars from \S\ref{s:phot.startrans}, although the spectroscopically
  confirmed quasars are \emph{plotted} here using the quasar
  transformations (\S\ref{s:phot.qsotrans}); the resulting color
  difference of at most 0.1 accounts for the few quasars appearing
  redder than $\ub=-0.3$.}
\end{figure*}

Figure~\ref{f:simPG.UBB} shows the color-magnitude diagram of our SDSS
parent sample, defined as all objects with $\ub < -0.3$ and $B<17.6$
(3 $\sigma_B$ fainter than the faintest PG plate limit) found within
4\fdg33876 of a PG survey plate center.  This parent sample includes
546 objects which have been identified as quasars in the SDSS
spectroscopic survey (green circles, Fig.~\ref{f:simPG.UBB}; the DR3
spectroscopic data set only covers a subset of the PG area, so that
future SDSS data releases will contribute additional spectroscopic
quasars in this area).  We account for the calibration difference
between PG and SDSS $B$ magnitudes by recalibrating the PG limiting
magnitudes using the transformation from PG to SDSS magnitudes derived
above (Equation~\ref{eq:Bfit}).  We then calculate the PG detection
probability for each object in our parent sample from the values of
the limiting magnitude of each plate, the $\ub$ cut, and the
respective PG photometric errors, assuming a Gaussian error
distribution.  By summing the detection probabilities for all objects
in a given color or magnitude bin, we directly obtain the expected
color and magnitude distributions for the PG survey.

\begin{figure*}
\epsscale{0.75}
\plotone{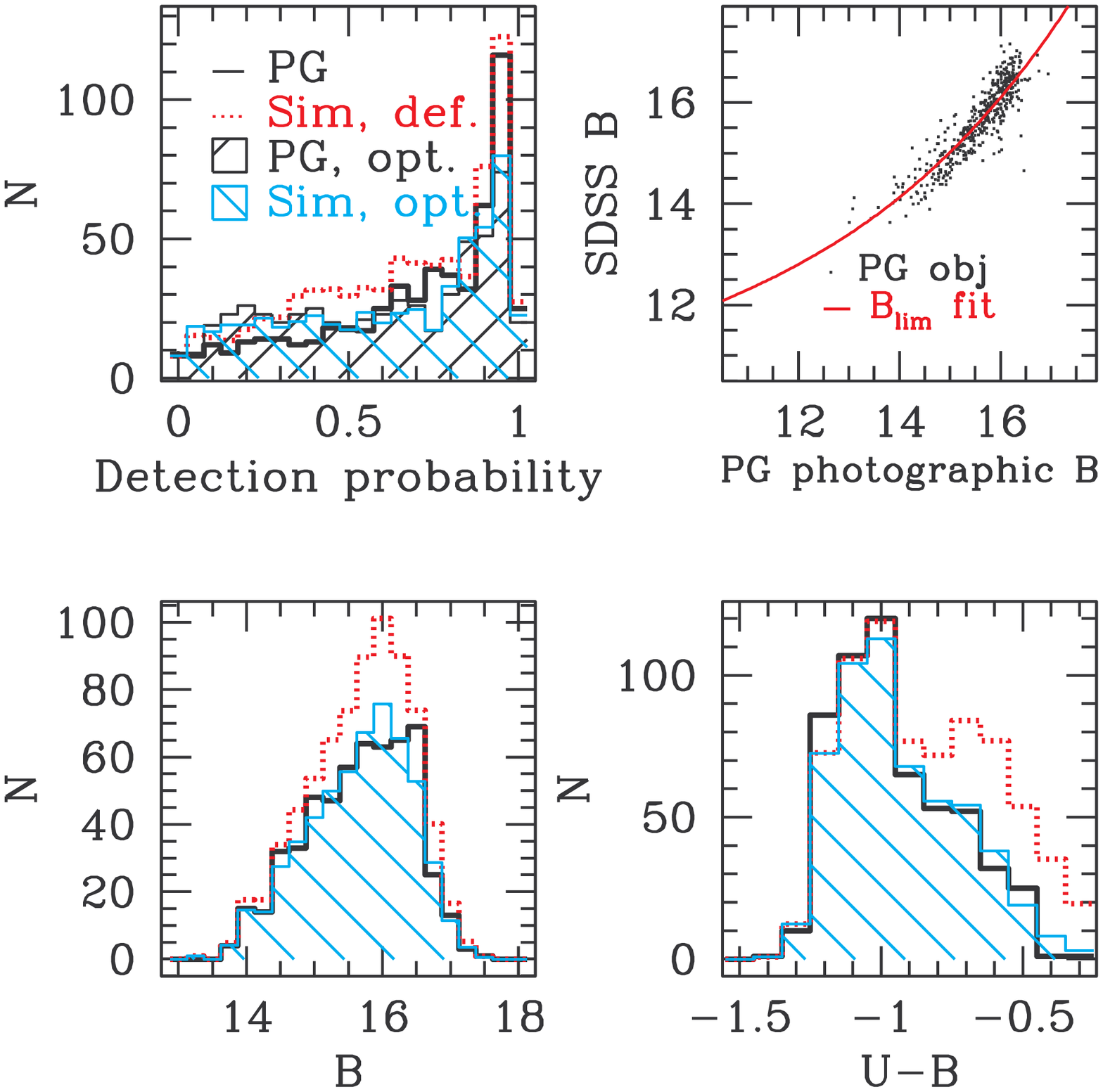}
\epsscale{1}
\caption{\label{f:simPG.Lhood}
Comparison of parameter distributions from actual PG and
SDSS-simulated PG survey.  We calculate a detection probability in the
PG survey for every SDSS object in Figure~\ref{f:simPG.UBB}, based on
the object's magnitude and color, each PG plate's limiting magnitude
after recalibration using the fit from Equation~\ref{eq:Bfit} as shown
in the upper right-hand panel, and two sets of limiting \ub, color and
magnitude errors: a ``default'' set with parameters $\sigma_B = 0.34,
\sUB = 0.36, (\ub)_\mathrm{lim}=-0.46$ as given by \citetalias{PG}
(thick solid and dotted red histograms) and an ``optimal'' set
$\sigma_B = 0.34, \sUB = 0.24, (\ub)_\mathrm{lim}=-0.71$ giving much
better agreement between the observed and simulated distributions
(black and cyan shaded histograms).  With the errors and color cut as
given by \citetalias{PG}, the PG survey as simulated from the SDSS
parent sample has many more objects with colors in the range $-1 < \ub
< -0.3$ and with apparent magnitudes in the range $15 < B < 16.5$ than
the actual PG survey.  To reduce the number of red objects in the
simulated PG and to match the slope of the $\ub$ histogram's cutoff
towards redder $\ub$ colors, it is necessary to adjust the color cut
to $\ub=-0.71$ and the $\ub$ error to $\sUB=0.24$ \citepalias[as
originally quoted by][]{BQS}.}
\end{figure*}

Figure~\ref{f:simPG.Lhood} shows the predicted distributions for $B$
magnitude, $\ub$ color, and detection probability (dotted red
histograms) and compares them to the distributions as observed in the
PG (black histograms).  The SDSS observations and the color cut and
photographic errors as given by \citetalias{PG} lead to an
overprediction of the number of PG objects with colors close to the
$\ub<-0.46$ cutoff and with magnitudes in the range $15 < B < 16.5$,
i.e., within a magnitude or so of the average limiting B magnitude.
These ``missing'' objects have PG detection probabilities near 50\%
based on SDSS photometry.  With hindsight, Figure~\ref{f:simPG.UBB}
already reveals that the PG is missing objects with $\ub$ colors close
to the cutoff: for a Gaussian error distribution, the detection
probability of an object close to the color cut is close to 50\%, but
PG detections nearly exclusively lie to the blue of the nominal color
cut of $-0.46$.  To reduce the number of red objects in the simulated
PG and to match the slope of the $\ub$ histogram's cutoff towards
redder $\ub$ colors, it is necessary to adjust the color cut to
$\ub=-0.71$ and the $\ub$ error to $\sUB=0.24$ \citepalias[as
originally quoted by][]{BQS}, resulting in the cyan shaded histograms
shown in Figure~\ref{f:simPG.Lhood}.  The offset between the corrected
$\ub$ cut and the value we find is comparable to the offset of about
0.2 magnitudes between SDSS and PG $\ub$ colors of quasars we
determined above (\S\ref{s:phot.sdss.colors} and
Figure~\ref{f:phot.UBdiff}).

There is a related test based on the number of main-sequence objects
scattered into the photometric sample. At a given $B$ magnitude, the
number of objects per unit $\ub$ color increases sharply into the main
sequence, redwards of $\ub=-0.3$. Therefore, the number of
main-sequence objects that are scattered into the PG photometric
sample and subsequently need to be removed in the spectroscopic
confirmation process is extremely sensitive to changes in the $\ub$
cut and error.  We repeat the detection probability calculation for
objects with $-0.3 < \ub < 0$ to obtain a lower limit on the average
number of main-sequence objects we expect to be scattered into the
photometric sample for a given combination of color and magnitude cuts
and errors.  We obtain a prediction of more than 17,000 main-sequence
objects being scattered into the PG photometric sample for $\ub<-0.44$
and $\sUB=0.38$, while the prediction for the revised values
$\ub<-0.71$ and $\sUB=0.24$ is roughly 500.  These values are for the
DR3-PG overlap area of roughly 3300 square degrees.  The actual number
of rejected main-sequence objects in the PG survey was 1125 over
10,668 square degrees, corresponding to roughly 300 over the overlap
area considered here, and consistent with our revised color cut and
photometric error.  This finding is confirmed independently by a
comparison of the sample of DA white dwarfs from the PG to samples
selected in other wavebands, which shows color incompleteness
beginning at $\ub=-1$ and increasing towards redder colors
\citep[][and Liebert 2004, \emph{priv.\ comm.}]{LBH04}.  This color
incompleteness is likely caused by the visual inspection process of
the measurement-machine selected UV excess candidates, which cut down
the number of candidates by a factor of 20 by removing objects that
appeared to have a neutral $\ub$ color on the photographic plates.

In summary, we find that the PG survey had an effective color cut $\ub
< -0.71 \pm 0.24$.  We now consider how different the BQS quasar
sample is from the BQS-like part of the SDSS quasar sample.

\section{Comparison of PG quasars to SDSS quasars}
\label{s:compPGSDSS}

The SDSS quasar survey reaches both to much fainter magnitude limits and
samples a larger portion of non-stellar color space than the BQS.  Our
comparison of these two quasar surveys has two aims:
\begin{enumerate}
\item To consider whether the BQS has any systematic
  incompleteness. To address this question, we compare the BQS quasars
  to those DR3 quasars passing the nominal PG criteria $B<16.16$ and
  $\ub<-0.44$ (we discuss this choice in detail in \S\ref{s:comp.opt}
  below.)
\item To consider the impact of the BQS selection criteria themselves,
  i.e., determine the selection effects present in the BQS. To address
  this question, we consider which part of the entire DR3 quasar
  catalog is selected by the BQS criteria.
\end{enumerate}

In order to obtain a meaningful comparison, we need to apply some cuts
to obtain comparable samples. First, SDSS quasar spectroscopic target
selection has a bright limit of $i>15$ to avoid fiber cross-talk and
saturation in the spectrograph.  We therefore apply the same bright
limit to SDSS matches of BQS quasars, leaving 39 BQS matches in the
PG-DR3 overlap area of approximately 3300 square degrees. Furthermore,
we restrict the SDSS spectroscopically identified quasars to $z<2.2$
quasars selected using the $ugri$ algorithm with faint limiting
magnitude $i<19.1$, since the $griz$ algorithm is explicitly aimed at
recovering quasars at $z>2.2$ to which the BQS UV excess cut is
insensitive.  The resulting sample consists of 30,975 quasars, drawn
from a subset of the DR3 area totaling 4188 square degrees.

\subsection{Comparison of optical properties}
\label{s:comp.opt}

\subsubsection{Selection of a BQS-like set of quasars from the SDSS
  quasar sample}
\label{s:comp.opt.bqslike}

Even though we established above (\S\ref{s:simPG}) that the PG
actually has an effective color cut $\ub<-0.71$, we select a
``BQS-like'' sample of SDSS quasars using the the intended limit
$\ub<-0.44$ because this is the value assumed by all previous
comparisons of other quasar surveys to the BQS.  Of the 30,975 SDSS
quasars remaining in the comparison sample, 26 satisfy the nominal BQS
criteria $\ub<-0.44$ and $B<B_\mathrm{eff}=16.16$.  We do not attempt
to determine the surface density of bright quasars in the SDSS here
since we have not yet completed the determination of the SDSS quasar
survey's selection function, which will appear elsewhere
\citetext{Richards et al. 2005, in preparation}.  

Furthermore, an object-by-object comparison between the BQS and
BQS-like SDSS quasars is not particularly meaningful. First, the large
photometric errors of the PG survey and the flux variability of
quasars mean that a reobservation of the PG area using the same
technology would not recover exactly the same set of objects. The BQS
quasars in the range $15.4 < B < 17.2$ only have an average detection
probability of 50\%.  Moreover, all quasars within $3\sigma$ of the
intended PG cuts in color and magnitude (i.e., with photographic $B <
B_{\mathrm{lim}} + 3\, \sigma_B$ and $(\ub) < (\ub)_\mathrm{lim} + 3\,
\sUB$) have a non-negligible probability of being included in the
sample, but their inclusion probability is only 10\% on
average. Therefore, it is very likely that an entirely different set
of objects will be scattered into repeat observations of the entire
BQS.  Variability on average decreases the re-detection probability
even more because quasars from a flux-limited sample on average become
fainter.  As a further consequence, the \emph{number} of objects
scattered into a BQS-like quasar sample will vary from reobservation
to reobservation. Using the PG parent sample constructed in
\S\ref{s:simPG}, we simulated 1000 realizations of the BQS, by
computing the detection probability of each of the 546 confirmed SDSS
quasars in the parent sample from the corrected PG limiting magnitudes
and the true limiting $\ub = -0.71$.  The number of quasars included
in the simulated BQS follows a Gaussian distribution with a mean of
27.9 and standard deviation of 4.4. In fact, 23 of the 546 SDSS
quasars considered here have also been detected by the PG.  Thus, the
number of actual BQS quasars in the overlap area between the PG and
the DR3 spectroscopic surveys is fully consistent with the range
observed in the Monte-Carlo simulations of the BQS.  This number is
different from the number of BQS objects for which SDSS photometry is
available (52 objects; \S\ref{s:sdss.match}), from the number of BQS
objects with SDSS photometry and $i>15$ (39 objects) and from the
number of ``BQS-like'' SDSS quasars with $i>15$, $B<16.16$ and $\ub <
-0.44$ (26 objects; see previous paragraph) because the comparison in
this section is limited to objects which have both an SDSS spectrum
and lie within 4\fdg33876 of the PG plate centers listed in Table 2 of
\citetalias{PG}, i.e., to a smaller area of sky. Further differences
between the number of SDSS and BQS quasars to the same flux limit
arise because of Eddington bias \citep{Edd13,Edd40}.  However, the
usual correction formulae \citep[see also][p. 162 ff., e.g.]{Pet97}
only give an \emph{average} correction, while the \emph{true} bias
depends on the details of the number counts at $B>B_{\mathrm{lim}}$
and the error distribution. In combination with the large errors of
photographic surveys, this makes it impossible to determine the
\emph{actual} Eddington bias with any degree of accuracy
\citep[see][]{Jef38,Lor04}.

Secondly, the limiting magnitudes of the PG survey plates vary
substantially (see Figure~\ref{f:intro.BQS.Blim}), and the surface
density of bright quasars is very low (most PG fields contribute 0 or
1 quasars to the BQS). Therefore, a comparison of the full BQS to a
survey with the same overall effective limiting magnitude, but a
different distribution of limiting magnitudes on the sky, will be
expected to result in many non-detections of BQS objects and
detections of non-BQS objects.

Hence, we prefer to compare the two surveys in a statistical sense, by
analyzing the distribution of BQS and BQS-like quasars in redshift,
magnitude and color space.  We consider each of the two-dimensional
projections of this three-dimensional parameter space in turn.  Using
matches between the FIRST radio survey \citep{BWR95} and SDSS data, we
also consider the radio properties.

\subsubsection{Magnitudes and colors as function of redshift}
\label{s:comp.opt.magz}

\begin{figure*}
\plottwo{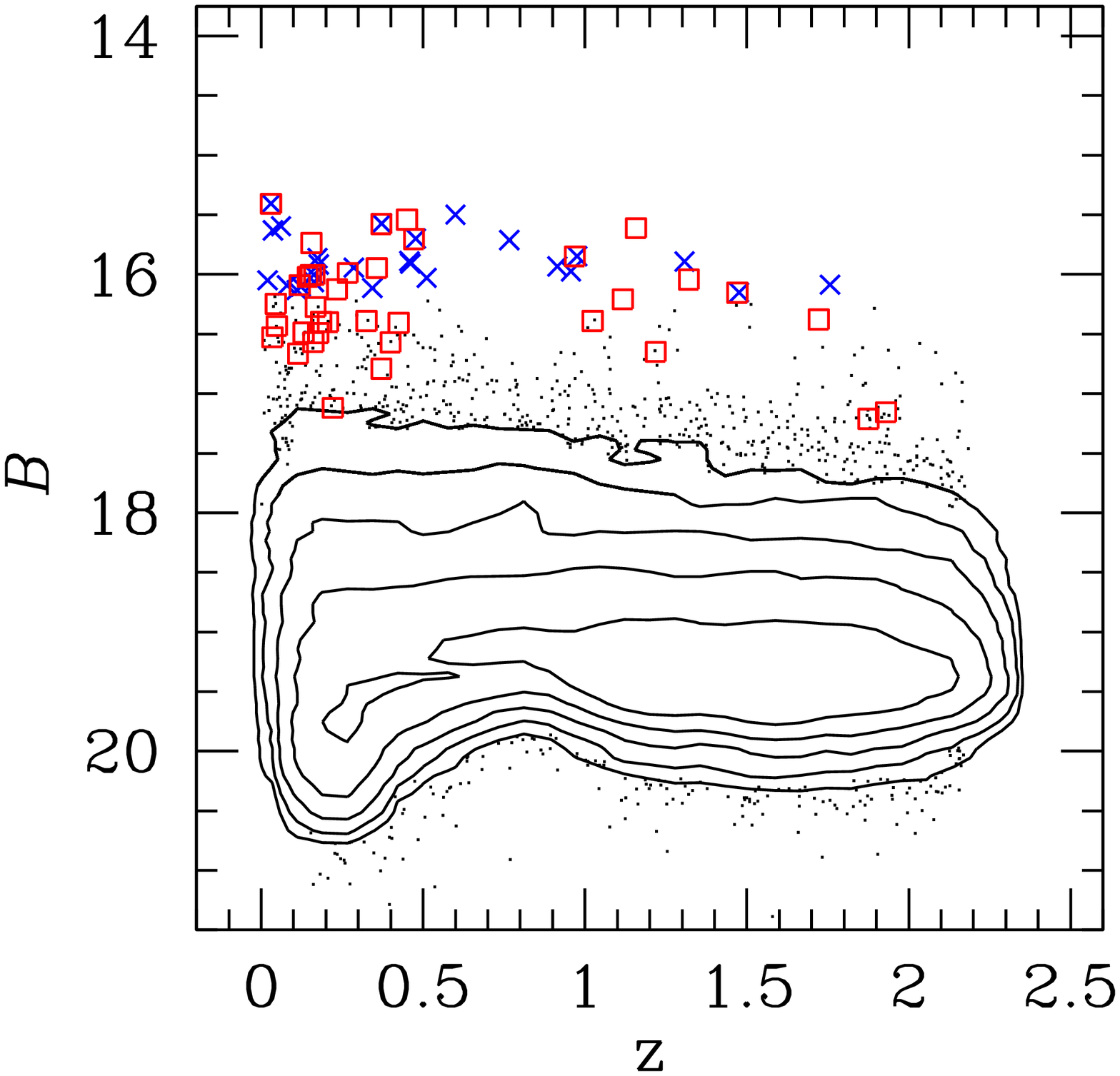}{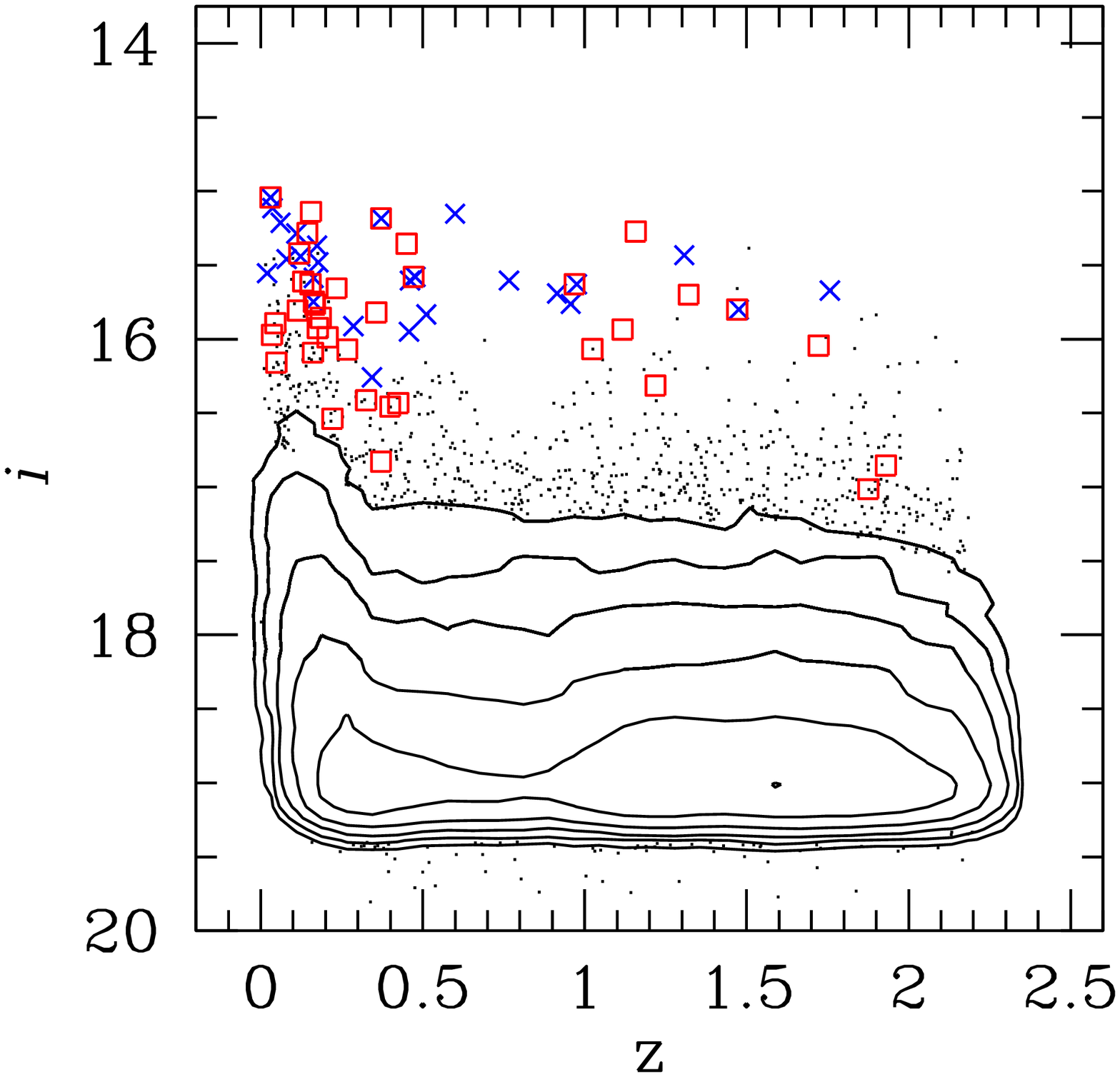}
\caption{\label{f:SDSS_Hubblediag}Hubble diagrams of SDSS DR3
  $ugri$-selected quasars at $z<2.2$ (dots and logarithmic contours
  increasing by a factor of 2), subset of DR3 quasars passing BQS
  criteria (crosses, blue in online edition), and BQS quasars in the
  DR3 area at $i>15$ (open squares, red in online edition).  Left,
  $B$-band; right, $i$-band.  The $B$-band diagram shows that the
  density of $B$-bright objects declines towards higher redshifts.
  The overdensity of bright $i$-band quasars at $z<0.3$ is a selection
  effect caused by the presence of the H$\alpha$ line in the $i$-band
  filters at these redshifts.  This also causes the corresponding
  overdensity of faint objects in the $B$-band diagram.}
\end{figure*}

It is instructive for the understanding of selection effects to
compare the properties of BQS quasars as well as BQS-like SDSS quasars
to the entire set of SDSS quasars.  We begin with the Hubble diagram,
the apparent magnitude plotted against redshift, in
Figure~\ref{f:SDSS_Hubblediag}.  Note that the number of bright
objects decreases with increasing redshift.  As expected from the
larger magnitude errors and the limiting magnitude varying from plate
to plate, the BQS quasars show some scatter about the effective
limiting magnitude $B_\mathrm{eff}$, while the BQS-like SDSS quasars
have a sharp magnitude cutoff.  Other than that, the BQS and the
BQS-like quasars have a similar distribution in the Hubble diagram.

\begin{figure*}
\plottwo{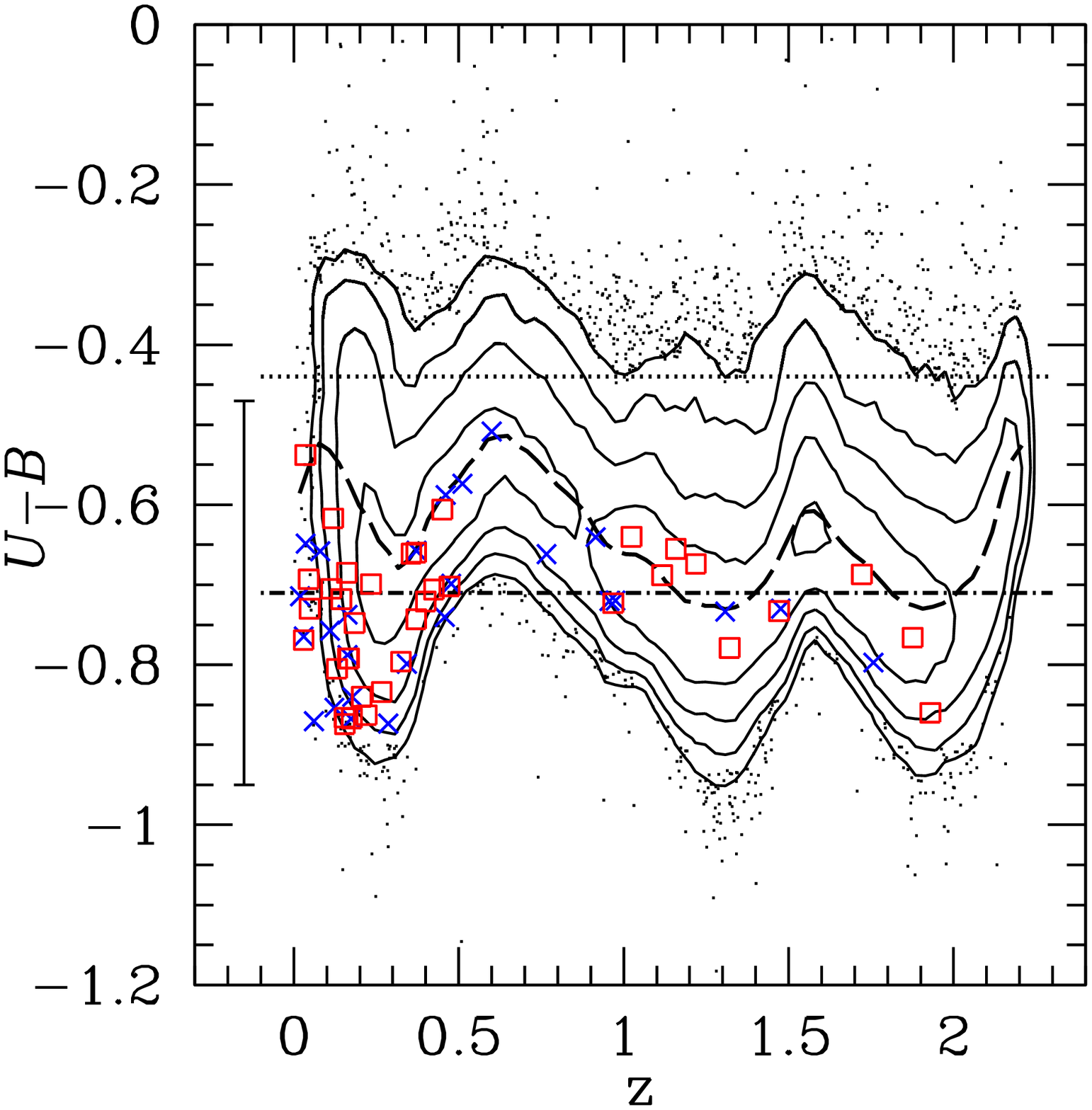}{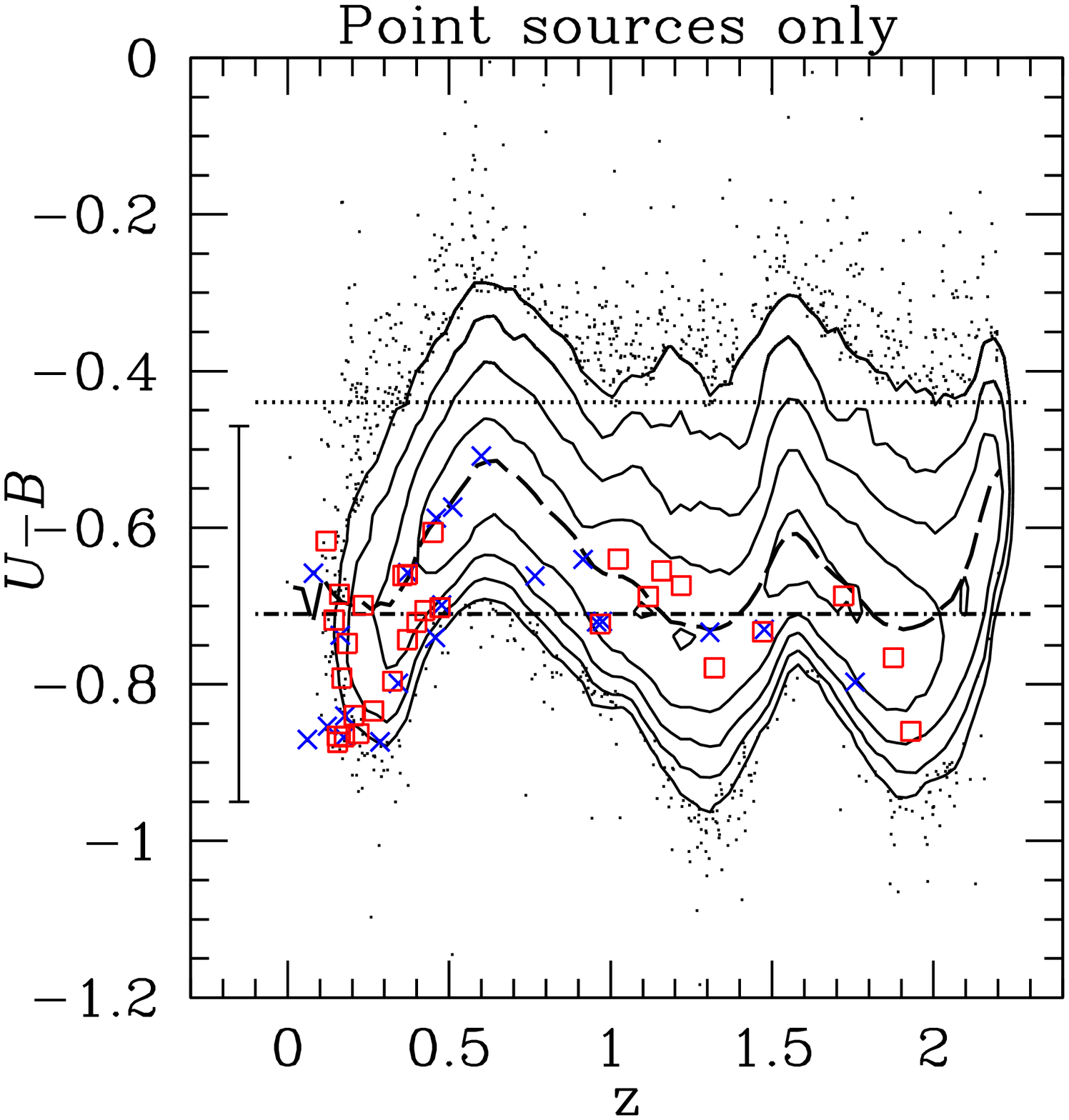}\\
\plottwo{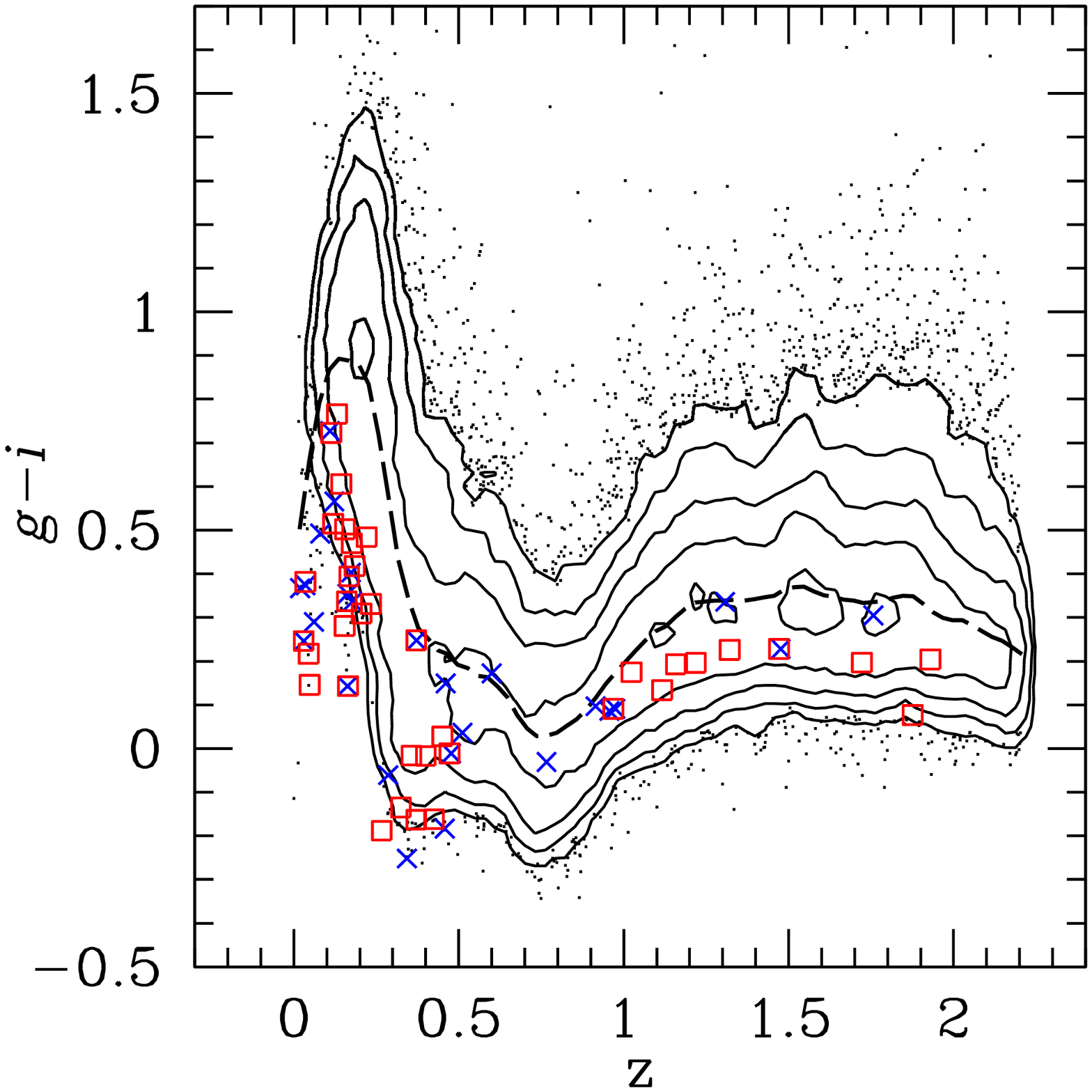}{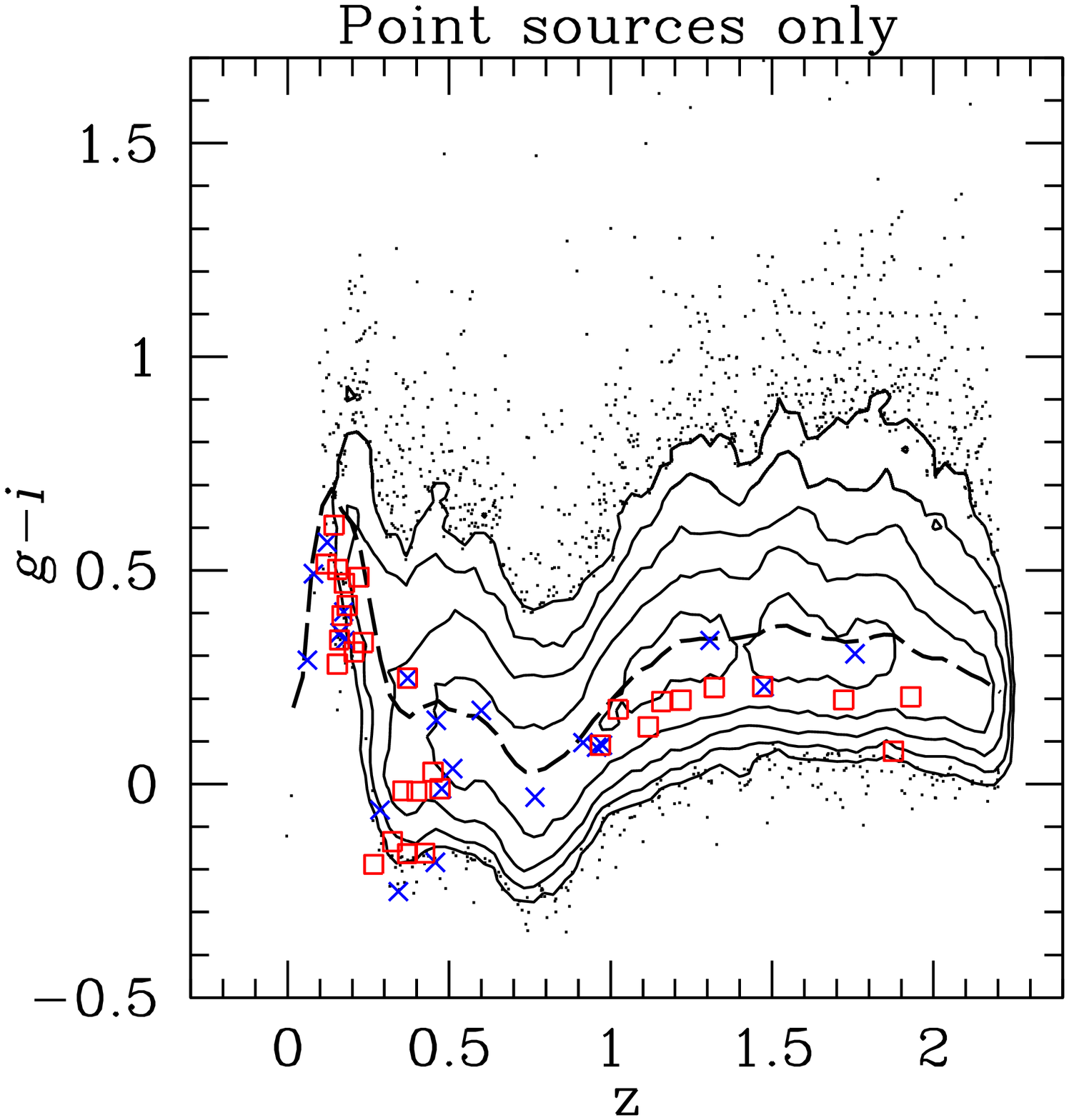}
\caption{\label{f:SDSS_colz}Colors of quasars as function of
redshift. Above, \ub; below, $g-i$. Left column, all SDSS
$ugri$-selected quasars at $z<19.1$. Right column, excluding sources
which are extended in SDSS photometry.  Symbols as in other figures.
The long-dashed line shows the median color in redshift bins of
0.01. The dotted line shows the nominal BQS color cut $\ub<-0.44$,
while the dot-dashed line shows the actual color cut $\ub<-0.71$ with
the error bar indicating the BQS \ub\ error $\sUB=0.24$ (see
Figure~\ref{f:simPG.Lhood}).}
\end{figure*}

Next, we consider the distribution of quasar colors against redshift
shown in Figure~\ref{f:SDSS_colz}. As noted before by several authors
\citep[e.g.][and references therein]{WCBea00,Ricea01}, there is a
systematic variation of the \ub\ color of quasars as a function of
redshift, caused by emission lines passing into and out of the two
filters.  As discussed in detail by \citet{WP85}, application of a UV
excess criterion in conjunction with this variation may lead to an
observed redshift distribution that is not representative of the true
redshift distribution.  But again, the distributions of both $u-g$ and
$g-i$ against redshift are very similar between the BQS and BQS-like
SDSS quasars.

In addition to the emission-line effect described here, quasar colors
can be reddened by the presence of host galaxy starlight, especially
at low luminosity (and therefore low redshift), where the host galaxy
is apparent in SDSS imaging.  Indeed, excluding extended quasars from
the color-redshift distribution (right-hand panels in
Figure~\ref{f:SDSS_colz}) bluens the median $\ub$ color at $z=0.2$ by
0.2 magnitudes; the median $g-i$ colors are affected even more
strongly. However, the BQS quasars are so luminous that the vast
majority appear as point sources in SDSS imaging; thus it would be
incorrect to compare colors of BQS quasars and the bulk of the SDSS
quasars at the same redshift.  Rather, we simply compare the
\emph{observed} \ub\ and $g-i$ colors of BQS and BQS-like SDSS quasars
at each redshift.  

A related concern is that the BQS criterion that quasar candidates
should have a ``dominant point-like appearance'' biases the BQS
against lower-luminosity quasars in higher-luminosity host galaxies.
Because of the correlation between the host's bulge mass and the black
hole mass, this will bias against sources with lower Eddington ratios;
together with correlations between luminosity or Eddington ratio and
continuum spectral slope, this may introduce a systematic color bias
\citetext{Ari Laor 2005, \emph{priv.\ comm.}}.  The importance of this
effect can be assessed by performing a decomposition of the SDSS
quasar spectra and colors into a galaxy and AGN component, which is,
however, beyond the scope of the present work.

In conclusion, the colors and magnitudes of quasars as function of
redshift are affected by the presence of emission lines and host
galaxy starlight.  These effects must be taken into account in the
analysis of redshift, magnitude and color distributions of quasars,
such as in the construction and interpretation of quasar luminosity
functions.

\subsubsection{Color-magnitude diagrams}
\label{s:comp.optt.cmd}

\begin{figure*}
\epsscale{.9}
\plottwo{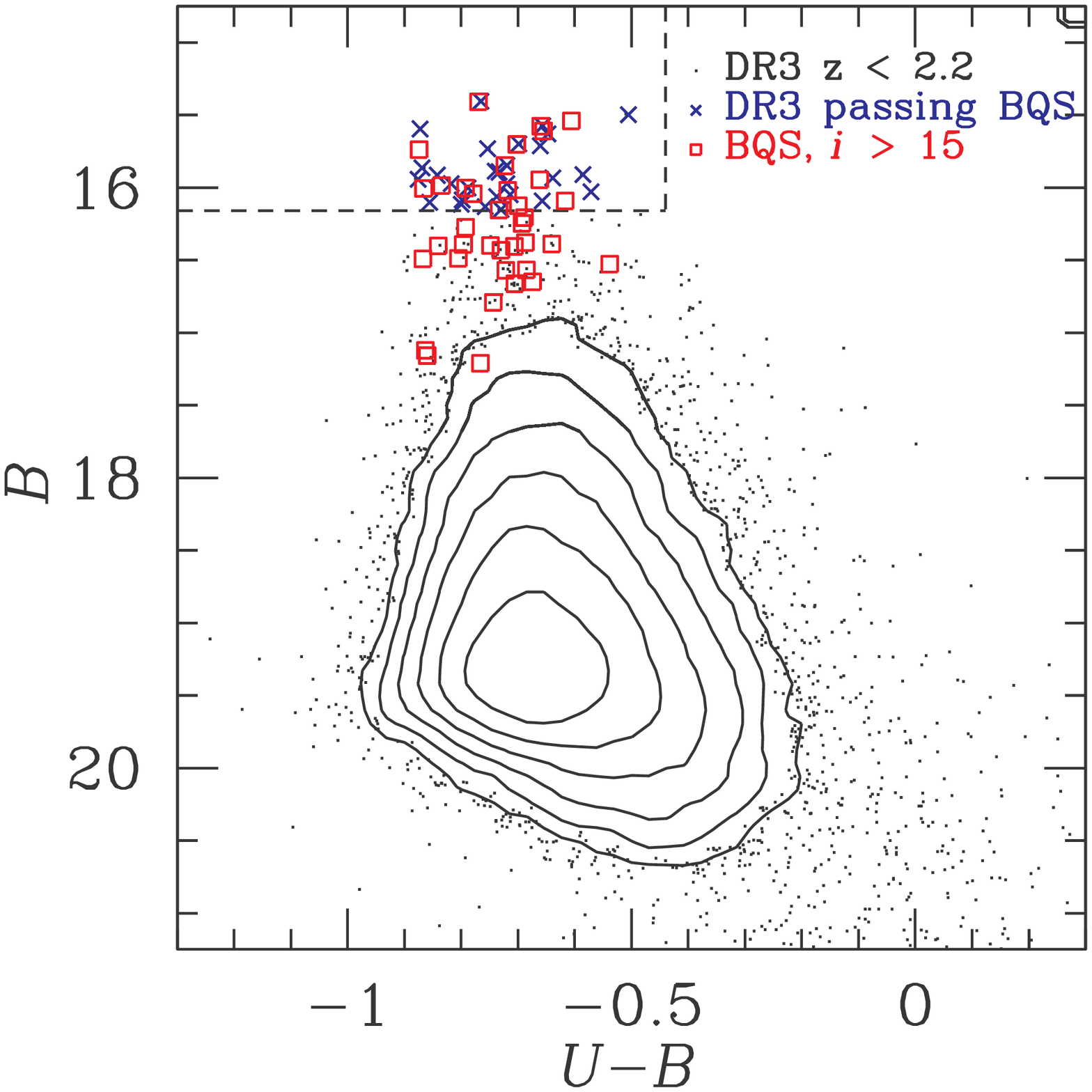}{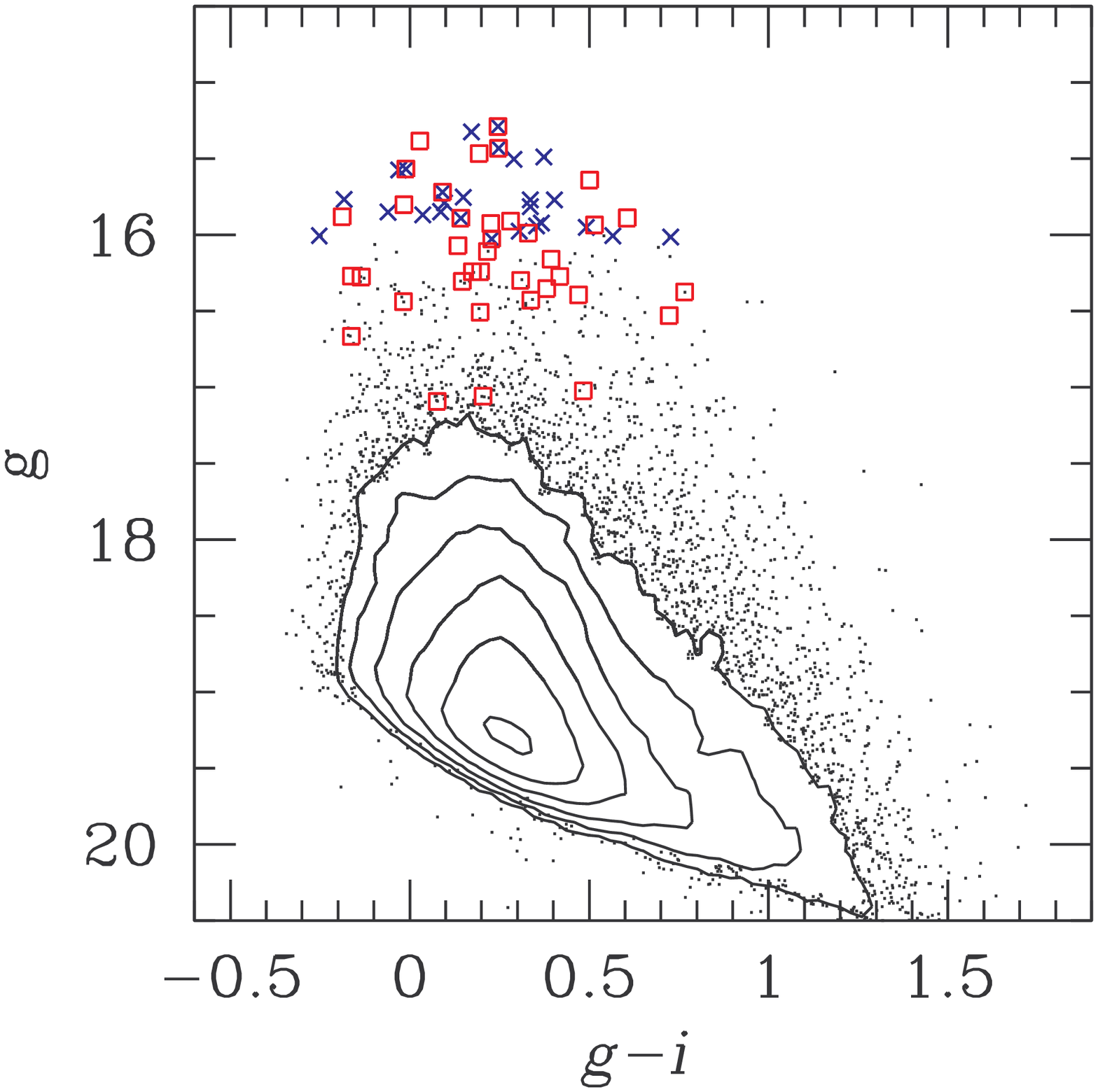}\\
\plottwo{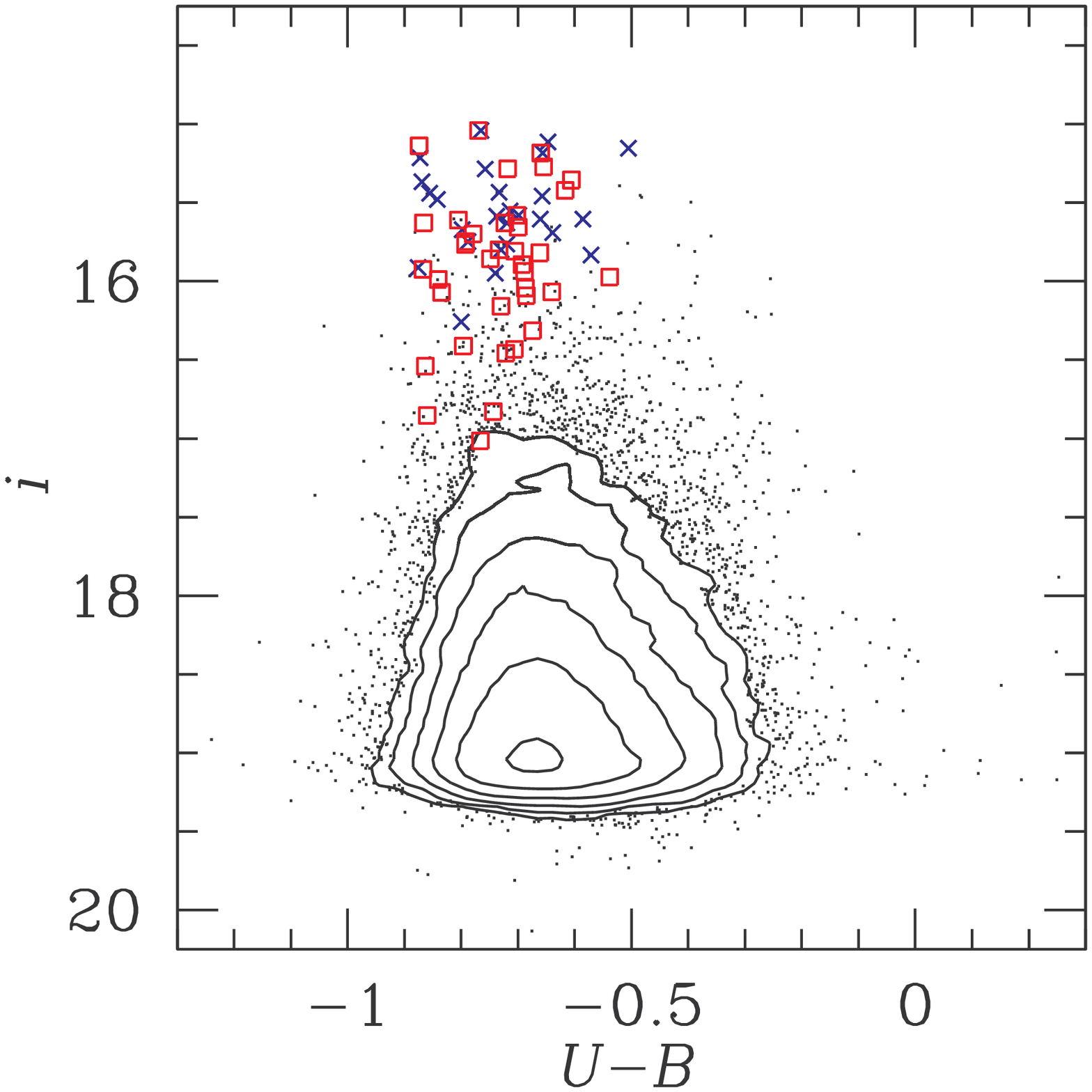}{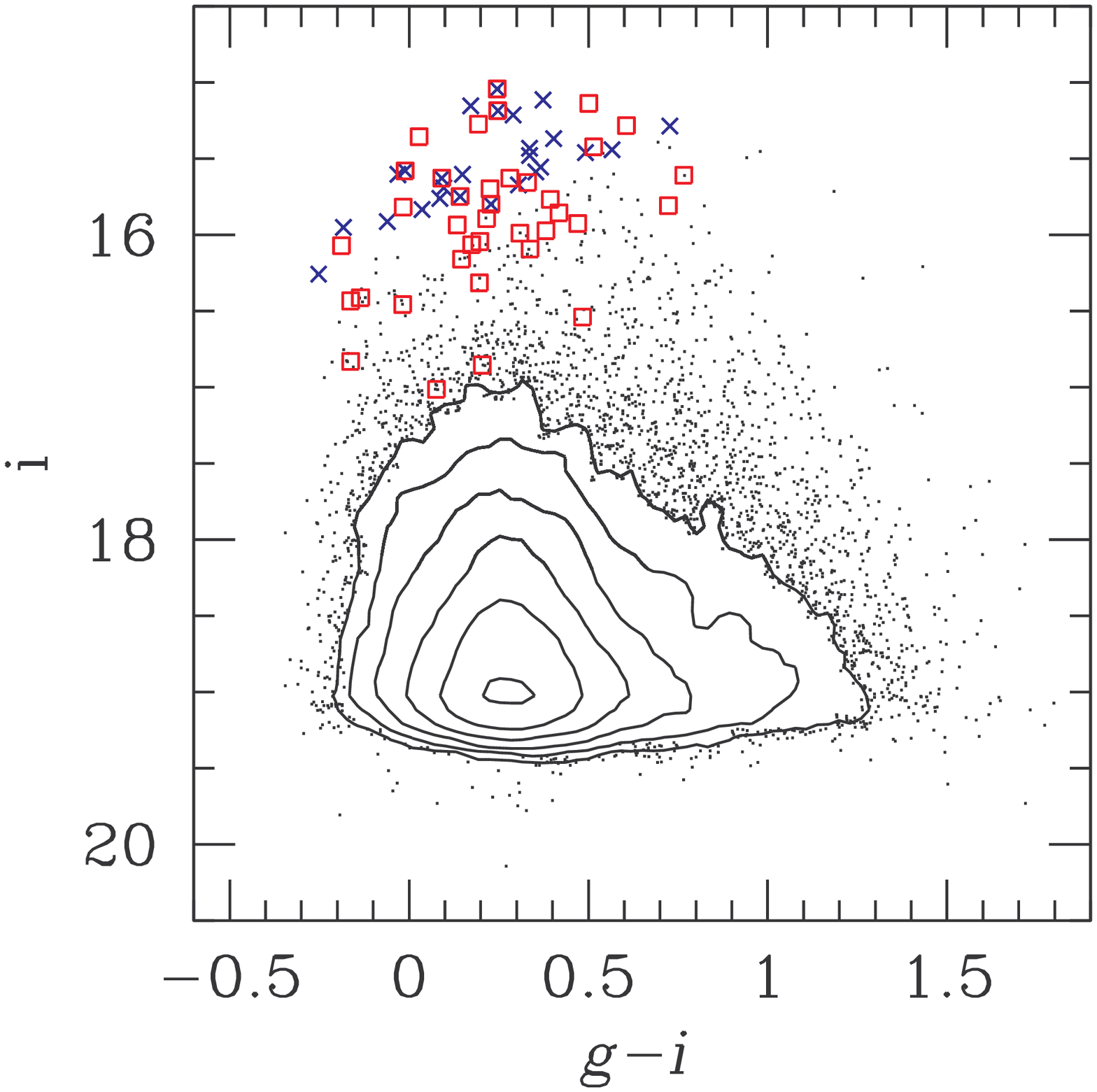}
\epsscale{1}
\caption{\label{f:SDSS_CMDs}Color-magnitude diagrams of SDSS and BQS
  quasars.  {\em Top left}, $B$ against \ub. {Top right}, $g$ against
  $g-i$. {\em Bottom left}, $i$ against \ub.  {\em Bottom right}, $i$
  against $g-i$ color (we use $g-i$ instead of $g-r$ or $r-i$ because
  colors derived from neighboring filters have strong features as a
  function of redshift where an emission line crosses from one filter
  to the next.). Contours and dots show the DR3 quasar sample, blue
  crosses are DR3 quasars passing the BQS limits shown by the dashed
  lines in the top left panel (i.e., $\ub < -0.44$ and $B < 16.16$),
  and open squares are SDSS data for the BQS $i>15$ quasars contained in
  the SDSS DR3. The correlation between \ub\ and $B$ at the faint
  cutoff is a selection effect caused by a combination of the
  H$\alpha$ passing through the $i$ filter and host galaxy
  contamination of quasar colors at low redshifts (see text).  The top
  left diagram shows that the BQS color cut does not remove any
  objects in addition to the BQS flux limit.  The other diagrams show
  that the more inclusive SDSS quasar candidate selection and the
  application of a flux limit in $i$-band include much redder quasars
  at bright $i$-band magnitudes than the BQS criteria, in particular
  compared to the BQS-like quasars selected from the SDSS.}
\end{figure*}

Figure~\ref{f:SDSS_CMDs} shows the color-magnitude diagrams (CMDs) of
the quasars at $z<2.2$ from the DR3 spectroscopic quasar sample and
the BQS quasars reobserved with the SDSS.  The SDSS sample is
flux-limited in $i$-band, as can be seen in the right-hand panel.  The
skewed cutoff to the $B$ against \ub\ distribution is caused by a
combination of two effects: one is that the H$\alpha$ line passes
through the $i$-band at low redshifts, boosting quasars that are faint
in $B$ above the $i$-band flux limit. The other effect is the increased
importance of red host-galaxy starlight for lower-luminosity quasars
we mentioned above.  Hence, quasars which are faint in $B$ are also
red in \ub.

The CMD shows clearly that the multicolor-selected SDSS quasars which
are sufficiently bright to pass the BQS limiting magnitude in the
$B$-band have a similar \ub\ color distribution to the BQS quasars
themselves.  Moreover, in the limit of small photometric errors, the
UV excess selection criterion does not remove any quasars from the BQS
sample that are not already removed by the $B$-band brightness cut (we
discuss the impact of the photometric errors below).  In other words,
the BQS is lacking red quasars not because of its UV excess criterion,
but because \emph{there are no red quasars which are bright in $B$}.

This can be understood by considering the shape of the \ub\
distribution at fixed $B$. The quasar density peaks at a constant
$\ub\approx -0.67$ at all $B$, and the dropoff in density towards both
redder and bluer colors is also independent of $B$ (the contours in
the flanks of the distribution are roughly parallel to each other down
to $B\approx20$, where the cutoff becomes skewed as discussed above).
Since the number counts of quasars fall rapidly with increasing
brightness, nearly all BQS quasars are drawn from near the mode of the
UV color distribution. To find significant numbers of both redder
\emph{and bluer} quasars than contained in the BQS, it is simply
necessary to sample fainter magnitudes (or a larger volume of the
Universe --- in fact, a much larger volume than is actually
available).  Thus, the BQS criteria do not produce any color bias when
considering only the UV colors of quasars that are bright in $B$-band.
However, the $i$ against \ub\ and $i$ against $g-i$ CMDs shows that
the BQS criteria do introduce a color bias against red quasars which
are bright in $i$-band.

The SDSS selects such red quasars simply because it applies the
brightness limit in the $i$-band.  The faintest (in $i$) BQS-like SDSS
quasar has $i=16.26$. There are only 26 DR3 quasars passing the BQS
criteria, but there are 109 DR3 quasars with $i<16.26$ (even after
rejecting extended sources, there are still 64 such quasars). Of the
109 sources, 96 have $\ub<-0.44$, while the remaining 13 redder
quasars have colors extending to $\ub=-0.26$ (5 of these are point
sources).  A Kolmogorov-Smirnov (KS) test comparing the $g-i$
distributions of the BQS quasars and of all SDSS quasars at $i<16.26$
rejects the hypothesis that they are indistinguishable at $>99\%$
confidence level; the $\ub$ distributions are formally just
indistinguishable (only 86\% confidence level for a rejection), but
the $\ub$ distribution of the $i<16.26$ SDSS quasars clearly extends
to redder colors than that of the BQS and BQS-like quasars.  Thus, the
BQS selection criteria lead to the omission of quasars with red
colors.  This bias is not driven primarily by the UV excess criterion,
but instead by application of the magnitude limit in the $B$-band.

\begin{figure}
\epsscale{0.45}
\plotone{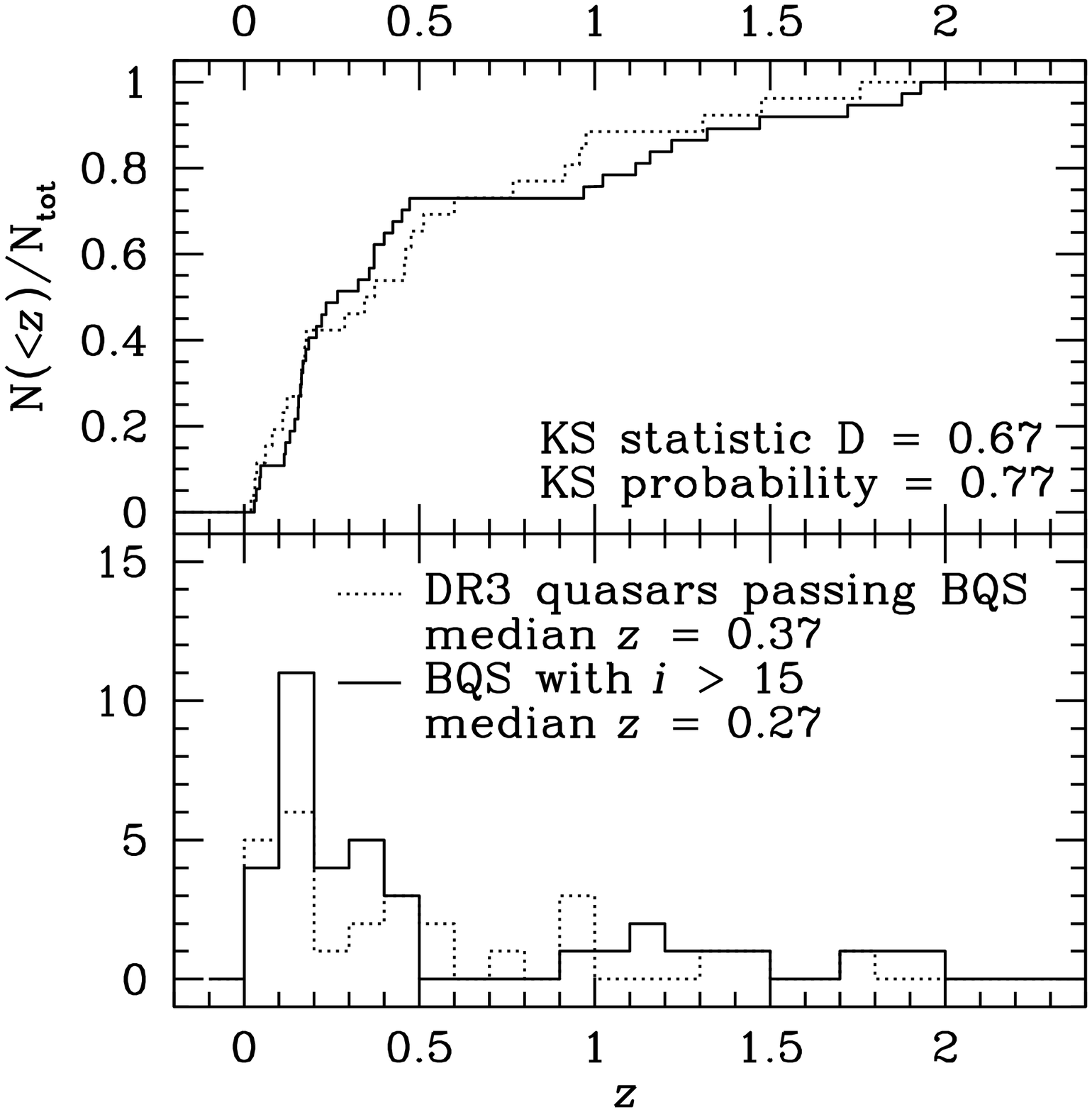}
\plotone{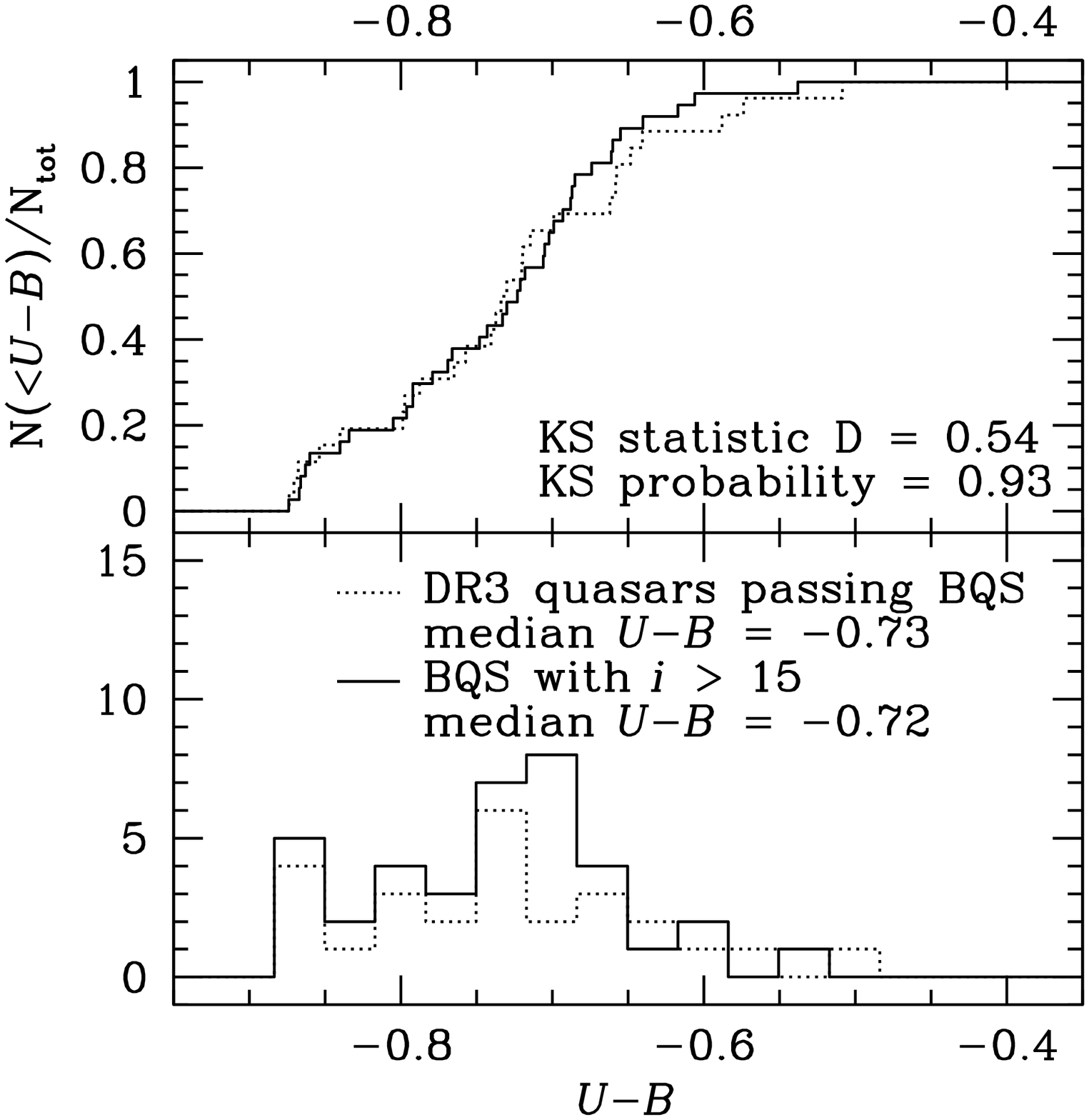}\\
\plotone{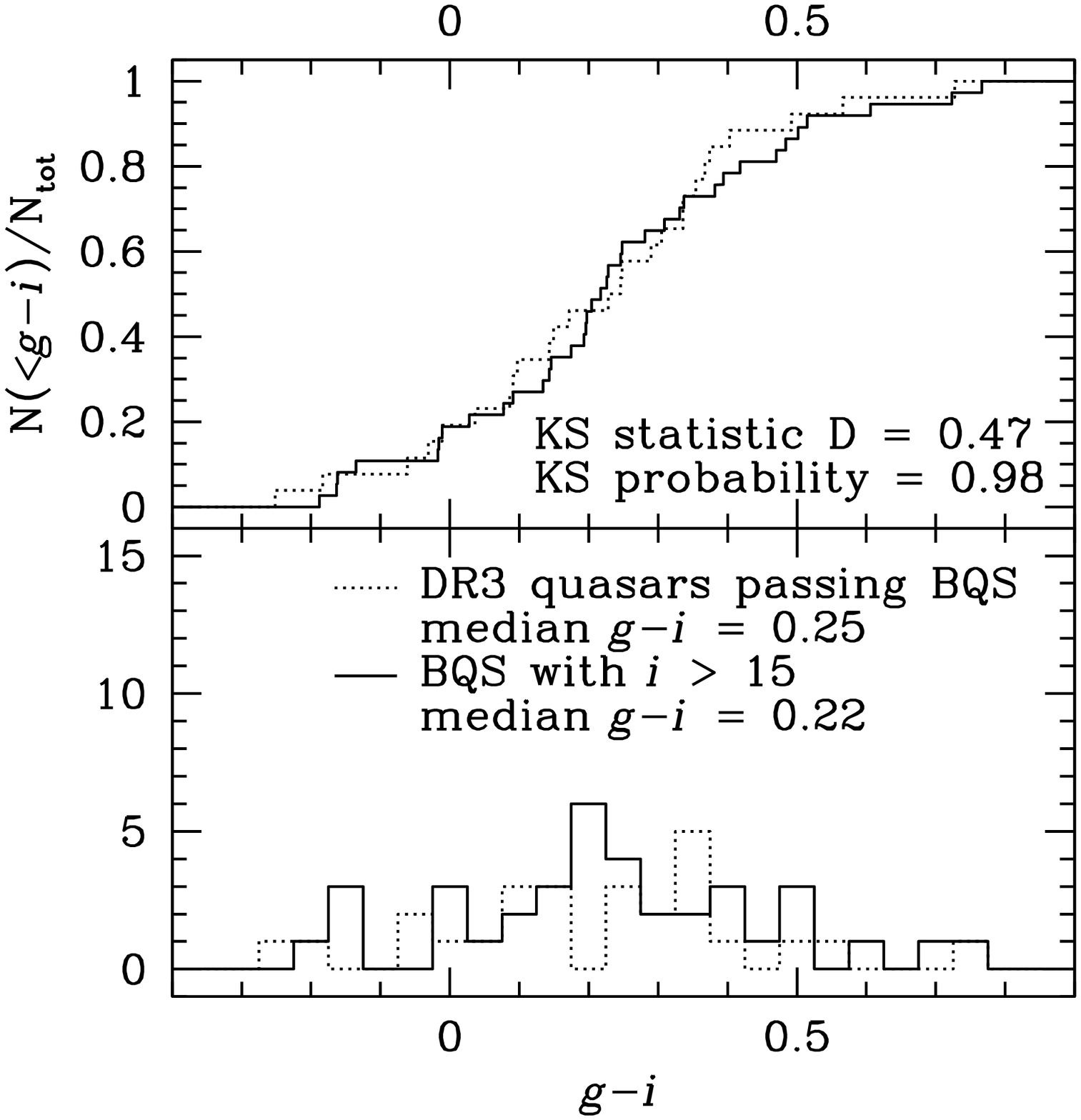}
\epsscale{1}
\caption{\label{f:SDSS_KStest}Comparison of redshift, \ub\ and $g-i$
cumulative and differential distribution of $i>15$ BQS quasars (dotted
lines) and DR3 quasars passing BQS criteria (solid lines).  All three
distributions are indistinguishable.}
\end{figure}

We now return to the comparison of the of BQS-like SDSS quasars and
actual BQS quasars. Figure~\ref{f:SDSS_KStest} compares the redshift
and color distributions of the 39 $i>15$ BQS objects inside the DR3
area to that of the 26 BQS-like SDSS quasars. The redshift
distributions are indistinguishable by the KS test, which returns a
very high probability for obtaining both observed distributions as
realizations of the same underlying distribution.  In fact, the
redshift distribution of the BQS quasars is indistinguishable at 95\%
confidence level from that of DR3 quasars with $\ub<-0.44$ down to
$B=17.0$.  The distributions of \ub\ and $g-i$ are similarly
indistinguishable between the BQS and the BQS-like SDSS quasars.  Even
though the BQS has relatively large color and magnitude errors and a
number of quasars are scattered into and out of the sample, the
optical properties of BQS quasars are not statistically
distinguishable from BQS-like SDSS quasars.  Thus, based on our
comparison sample, the BQS is a representative survey of bright, blue
quasars.

To summarize: Of the SDSS quasars above some limiting $i$, only the
bluest ones are bright enough in $B$ to pass the BQS $B$-band
cut. Therefore, the BQS flux cut excludes most SDSS quasars which have
comparable $i$-band magnitudes as the $i$-faintest BQS objects. Only
an additional 10\% are redder than the BQS \emph{color} cut.  To
recover the red end of the quasar color distribution, it would be
necessary to extend the search to fainter $B$ magnitudes \emph{in
addition to} relaxing the UV color excess criterion.

\subsection{Radio properties}
\label{s:comp.radio}

\subsubsection{Continuum properties}
\label{s:comp.radio.cont}

\begin{figure*}
\plottwo{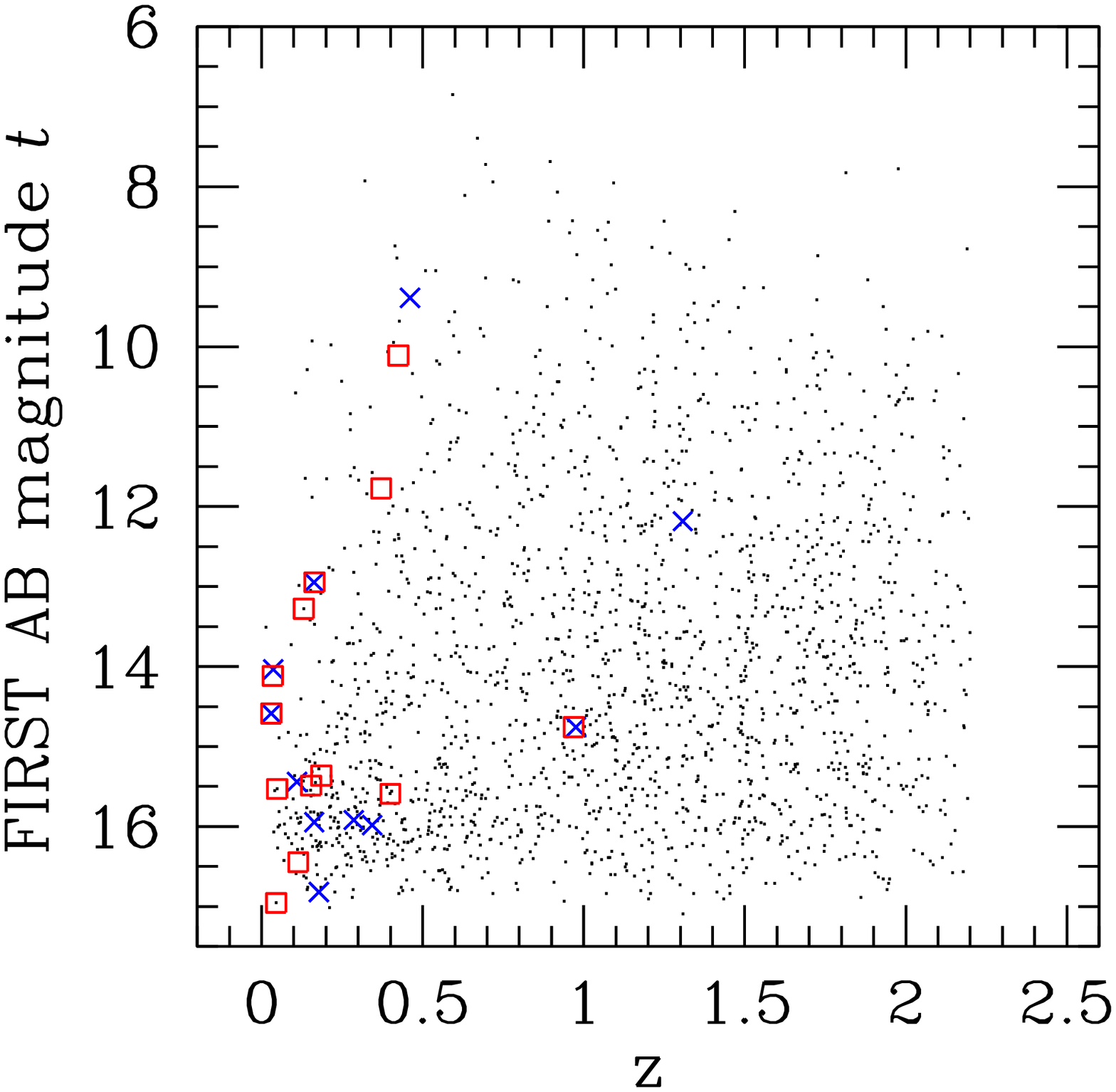}{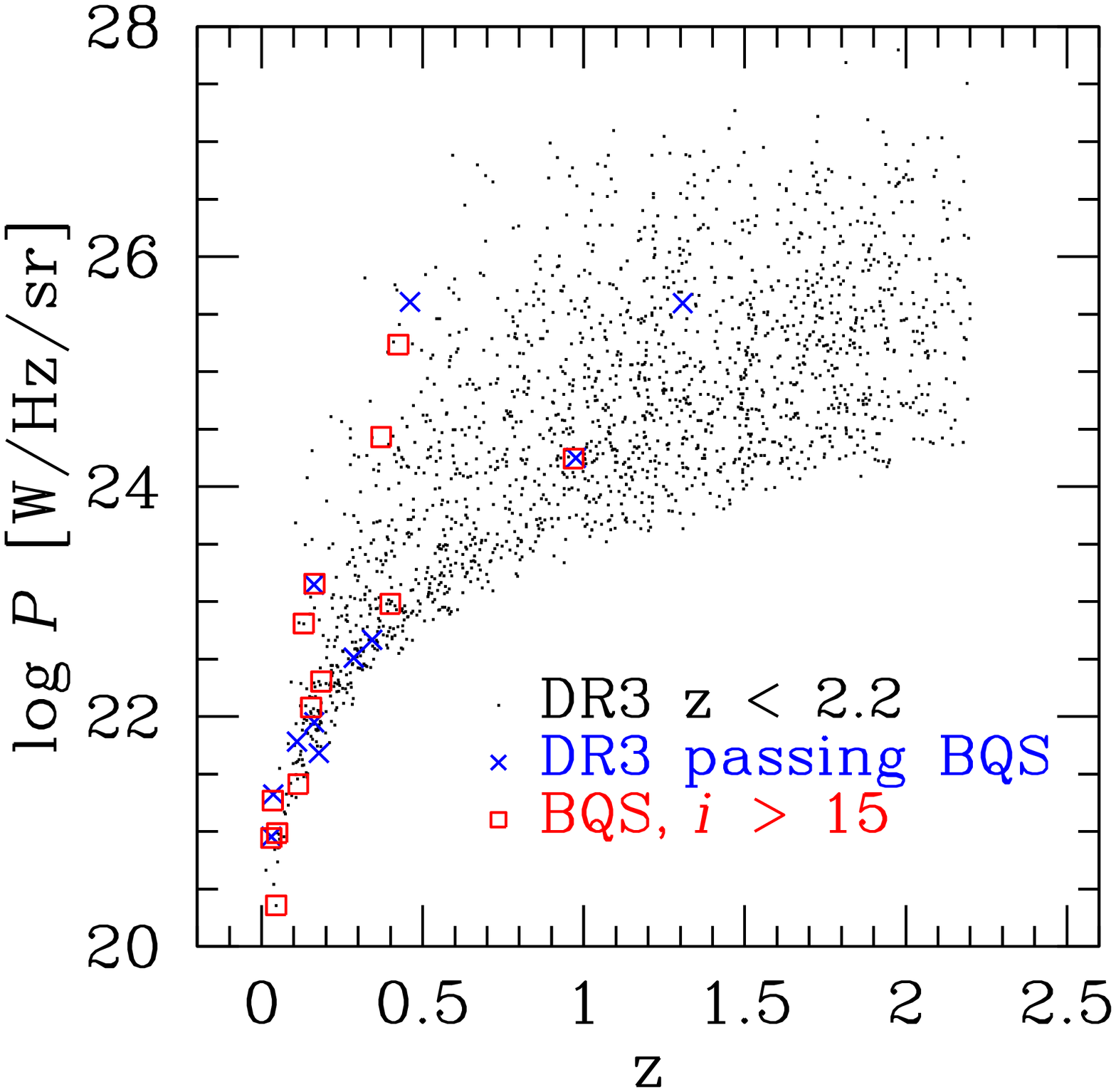}\\
\plottwo{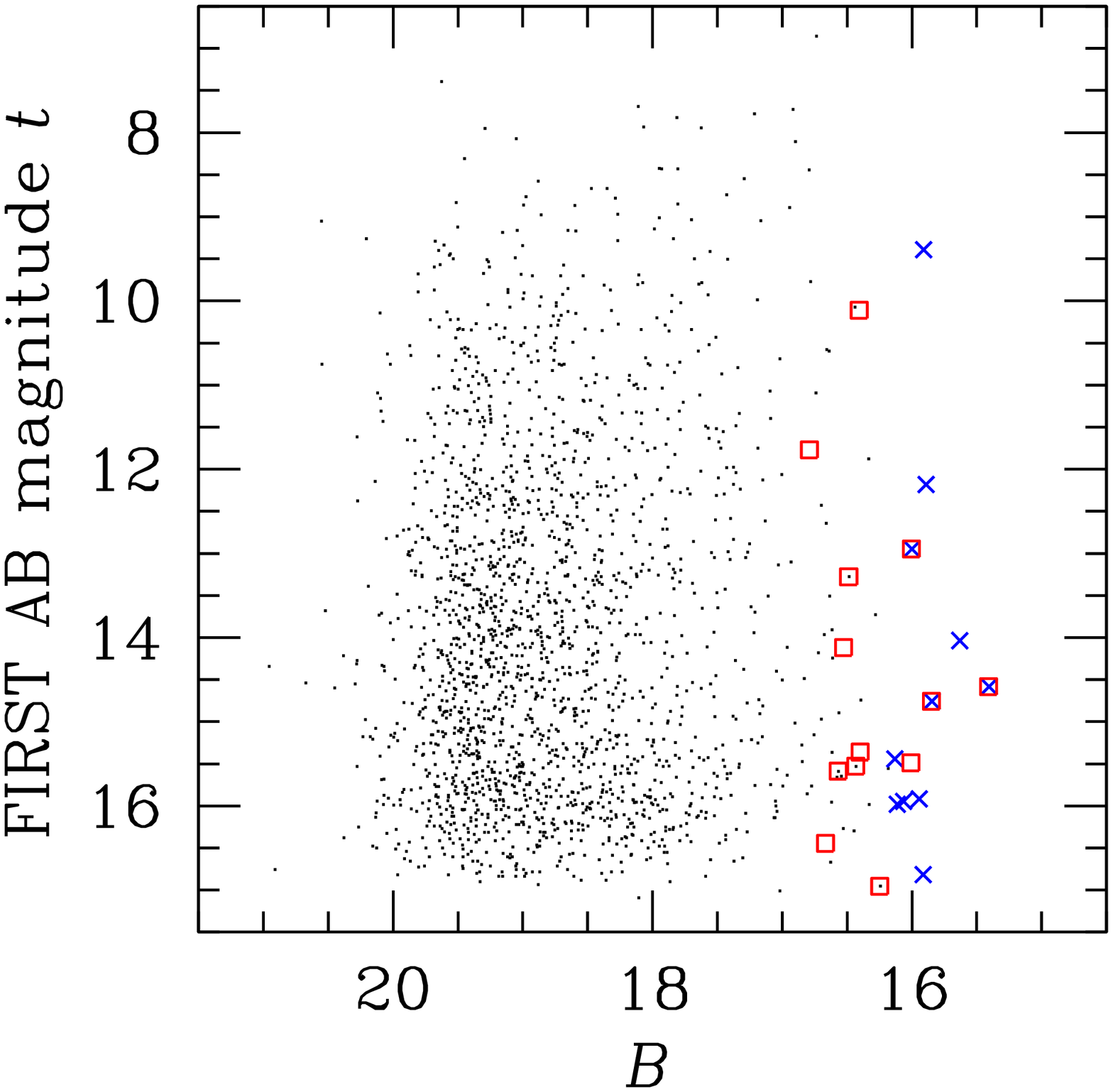}{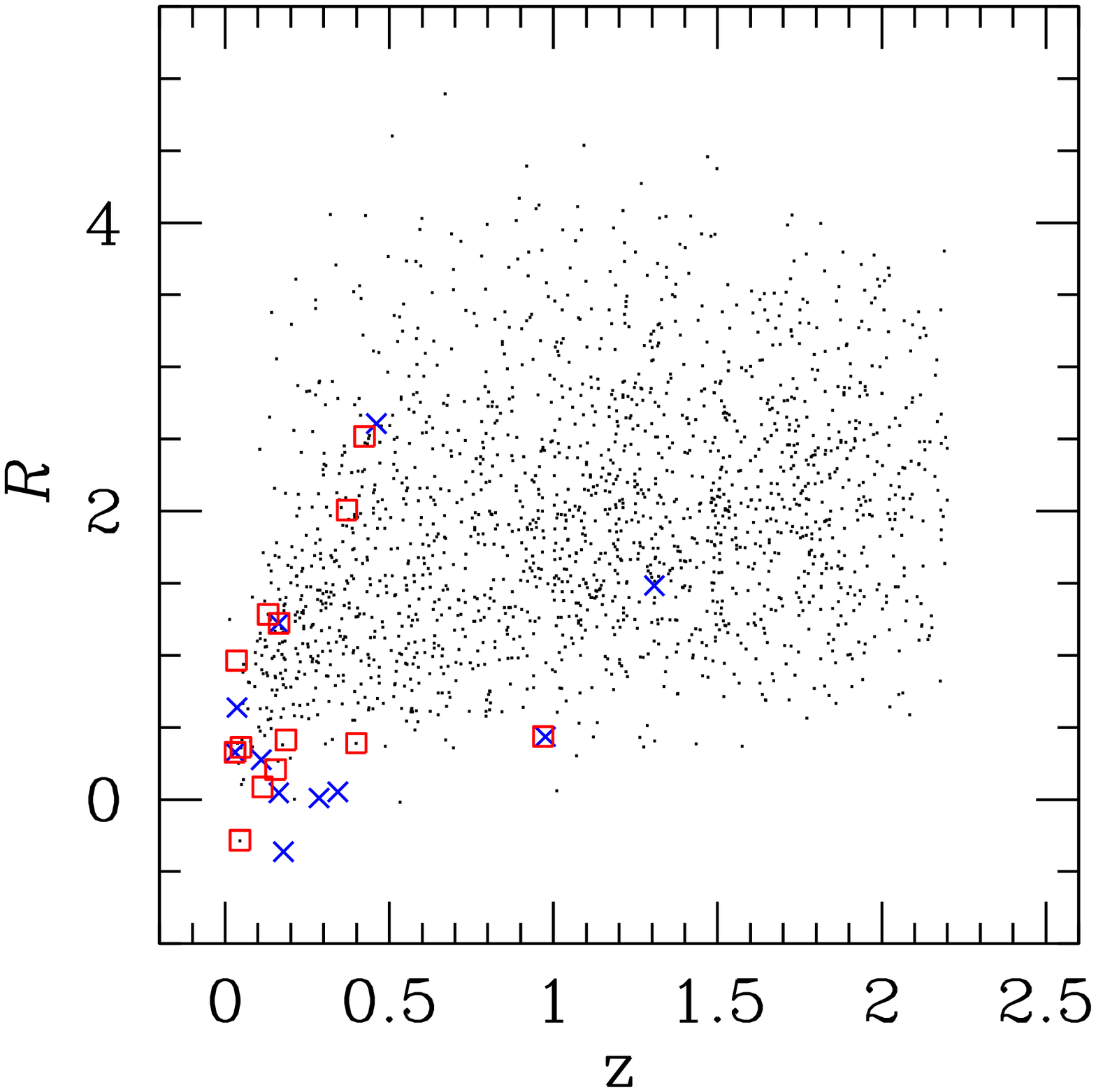}
\caption{\label{f:radio_z}FIRST properties of SDSS and BQS quasars,
showing integrated radio flux expressed as AB magnitude $t$ (top
left), core radio power $P$ (top right), radio-optical ratio $R = 0.4(B-t)$
(without any $K$-correction, bottom right) against redshift, as well
as radio AB magnitude against $B$ magnitude (bottom left). Symbols as
in previous figures. There are no obvious gaps in any of the
distributions, which are, however, subject to all selection effects of
both the SDSS/BQS and the FIRST surveys. The $R$ distribution is most
severely affected because objects below the flux limit of either
survey can have any value of $R$.  The concentration of low-$R$ points
at low $z$ is caused by the increasing number of sources which are
bright in $B$ towards low redshift (compare
Figure~\ref{f:SDSS_Hubblediag}).}
\end{figure*}

We next consider the radio properties of the quasars using matches to
SDSS data from the FIRST survey.  The FIRST survey itself has a flux
limit (1 mJy at 1.4 GHz) which introduces selection effects in
addition to those from the quasar surveys.  Use of the FIRST data is
preferable over the use of dedicated radio observations of the BQS
because it provides a homogeneous data set for comparing the radio
properties of the SDSS quasars to those of the BQS quasars.  Since
various claims of a radio-loud/radio-quiet bimodality have been made
based on the radio flux \citep{KSSea89}, radio/optical ratio
\citep[e.g.,][]{SHPea80,IMKea02} and the radio power $P$
\citep[e.g.,][]{PML86,MPM90,MRS93}, we begin by considering these
three quantities as a function of redshift.  We use the integrated
FIRST flux of all objects and \citep[following][]{IMKea02} define a
FIRST AB magnitude
\begin{displaymath}
t = -2.5 \log \frac{f_\mathrm{FIRST}}{3631 \mathrm{Jy}}
\end{displaymath}
and the logarithmic radio-optical ratio (without any $K$-correction)
\begin{equation}
  R = 0.4 (B-t).
\label{eq:R}
\end{equation}
The FIRST limiting flux of 1\,mJy corresponds to $t=16.40$. 

\begin{figure}
\epsscale{0.45}
\plotone{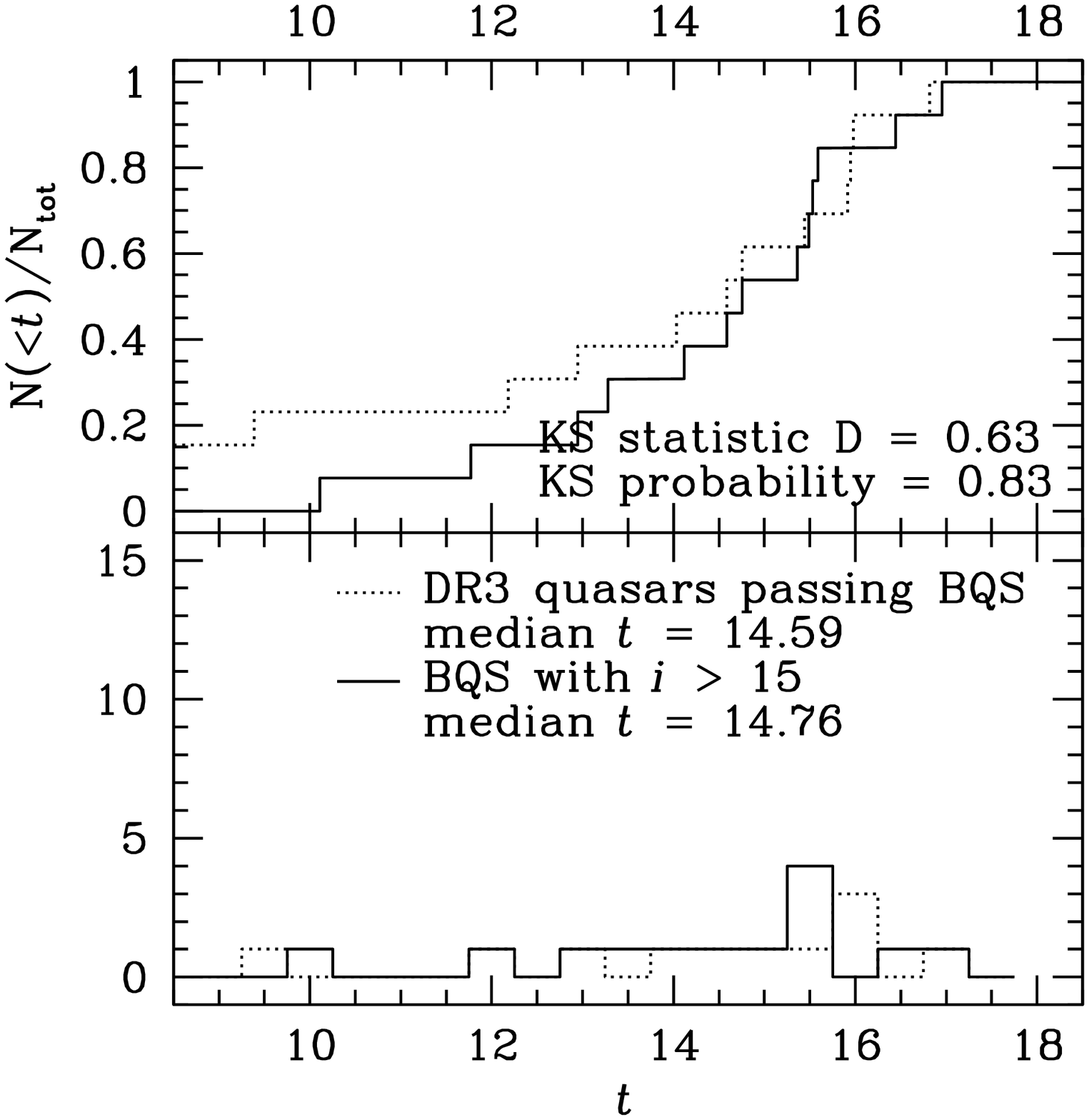}
\plotone{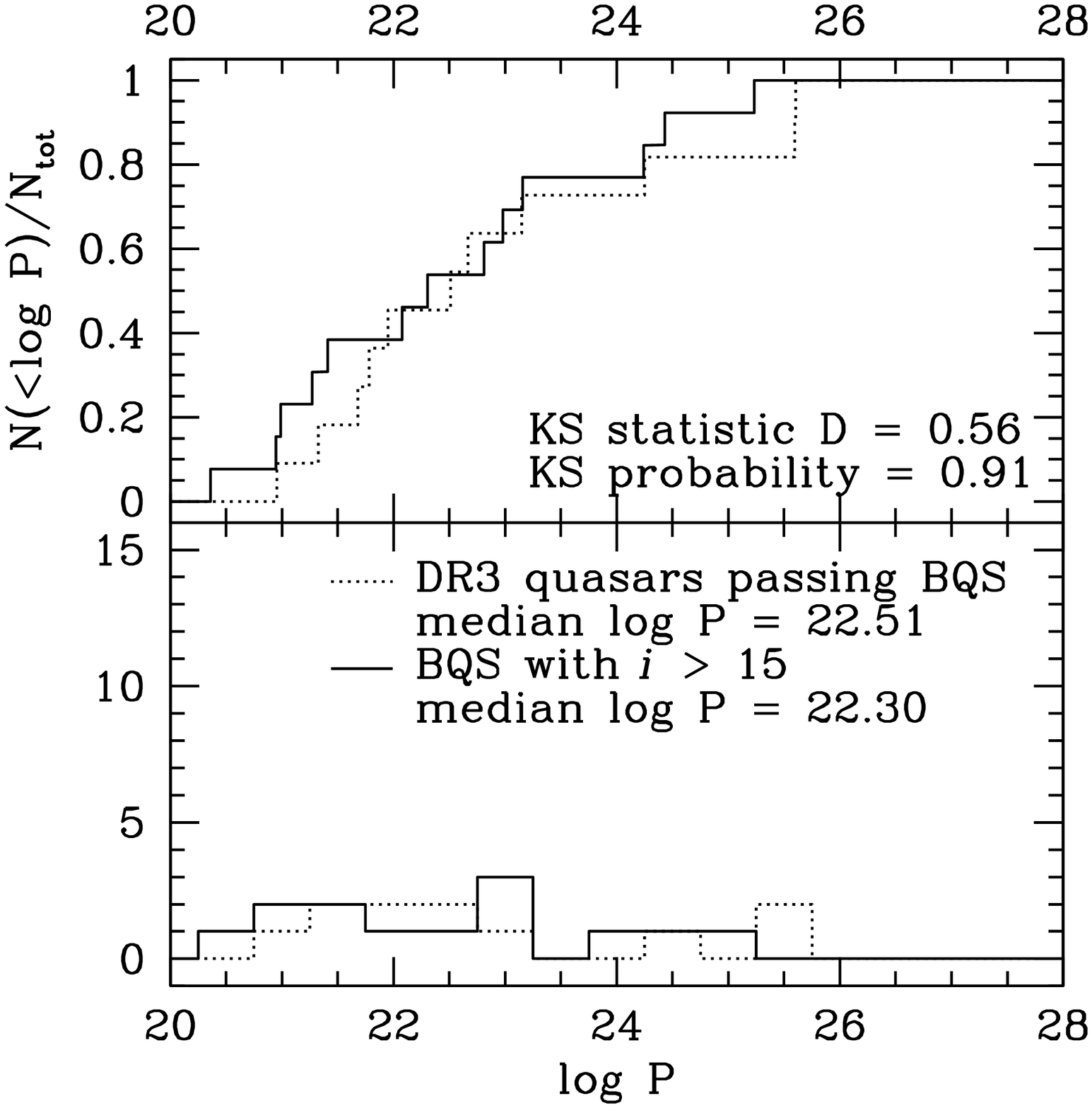}\\
\plotone{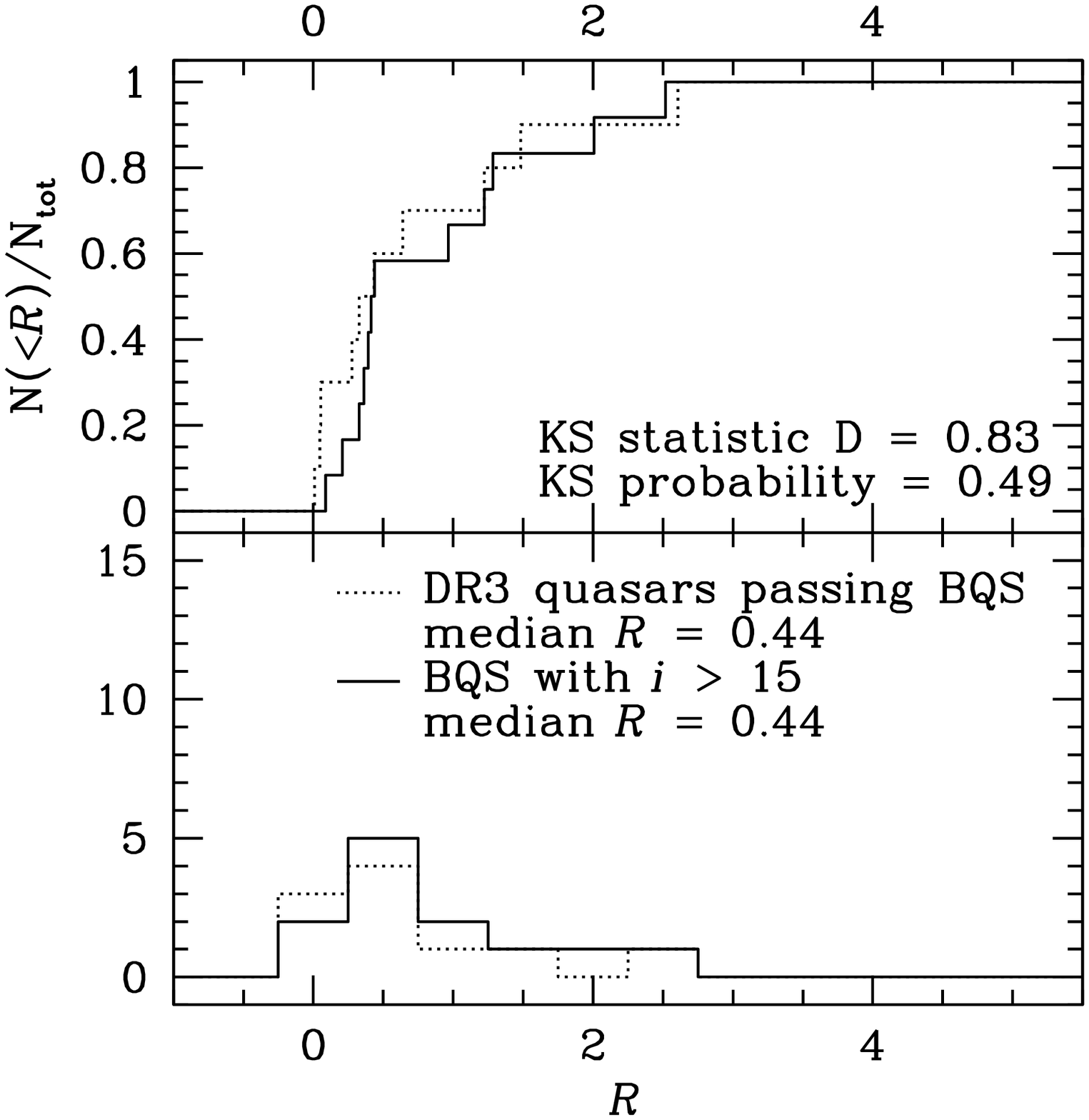}
\epsscale{1}
\caption{\label{f:radioKS}Comparison of the radio properties of $i>15$
BQS objects (dotted lines) and DR3 quasars passing BQS criteria (solid
lines).  The distributions in FIRST AB magnitude $t$, radio
power $P$ and radio-optical ratio $R = 0.4(t-B)$ are indistinguishable
between the two sets of quasars.}
\end{figure}

Figure~\ref{f:radio_z} shows the distributions of radio flux, power,
and radio-optical ratio for DR3 and BQS quasars.  There are no obvious
gaps in any of the distributions which would reveal a strong
bimodality in the radio properties of optically selected quasars.
Note, however, that the distributions are influenced by the selection
limits in \emph{both} the optical and radio surveys. For example,
objects with small value of $R$ are concentrated at low redshift
because the lowest $R$ is obtained for the objects which are brightest
in $B$, which are concentrated at low redshift.
Figure~\ref{f:radioKS} compares the distributions of $t$, $\log P$,
and $R$ for the two sets of quasars. In all three cases, the
distributions are again statistically indistinguishable, with KS
probabilities of 0.83, 0.91, and 0.49 for obtaining the observed $t$,
$\log P$, and $R$ distributions, respectively, from the same
underlying distribution.  Thus, we find no evidence for any radio
dependence of the BQS incompleteness.

We noted above that the $g-i$ distribution of the 26 BQS-like SDSS
quasars is different from that of all 109 SDSS quasars with $i<16.26$,
the faintest $i$ magnitude of the BQS-like quasars.  However,
according to KS-tests, the radio properties of these two samples are
entirely indistinguishable.  We will return to this point in the next
section, where we briefly consider the relation between emission-line
properties and radio emission of objects in our BQS-like sample.

\subsubsection{Possible BQS biases with respect to radio properties}
\label{s:comp.radio.bias}

As noted in the introduction, \citet{MRS93} stressed ``the rarely
appreciated fact that, at $z\sim 0.3$, about 50 per cent of the
brightest optically selected quasars \ldots are steep-spectrum,
radio-loud objects'' (these authors use the [OIII]~5007 narrow-line
luminosity to define ``brightest'', i.e., in fact they appear to mean
``most luminous'', and use an emission-line quantity rather than a
continuum measure such as the continuum flux at a fixed rest-frame
wavelength or the $B$-band magnitude).  Their footnote cautioning that
this number might be spuriously high if the BQS incompleteness favored
the inclusion of radio-loud objects has been interpreted as
\emph{suggesting} such a radio-dependent incompleteness --- however,
these interpretations have not been supported by any additional
evidence.  In fact, we determined in the previous section that the BQS
incompleteness appears to be random with respect to radio properties,
given that BQS-like DR3 quasars have indistinguishable distributions
of FIRST flux, luminosity, and radio-to-optical flux ratio.  However,
if the radio and optical properties of all or some quasars are
correlated in some way, the BQS sample may not be a random sample of
quasars with respect to radio properties.

As a rough check of their finding, we computed $L_{\mathrm{[OIII]
    5007}}$ from the $\mathrm{[OIII] 5007}$ flux determined by the
SDSS pipeline and our custom spectral-line fitting routine
\citetext{Stoughton, Vanden Berk, \& Jester 2005, in preparation} for all
SDSS quasars with $z<0.5$ and $B<17$, where we used a fainter flux
limit to increase the sample size to 116 objects.  All objects which
\citet{MRS93} classified as ``classical double radio sources'' (their
category D), i.e., radio-loud objects, have $L_{\mathrm{[OIII] 5007}}>
5\times10^{35}$\,W.\footnote{\citet{MRS93} used a cosmology with
  $H_0=50$\,km/s/Mpc, $\Omega_m = 0$, and $\Omega_\Lambda=0$,
  resulting in luminosities about a factor 2 greater than those in the
  current standard cosmology, which we employed ($\Omega_{\mathrm{m}}
  = 0.3, \Omega_\Lambda = 0.7$, and $H_0 = 70$\,km/s/Mpc).}  In our
sample, there are 13 sources with $L_{\mathrm{[OIII] 5007}}>
5\times10^{35}$\,W. Of these, four have a detection of FIRST flux at
the optical position.  Inspection of FIRST and NVSS \citep{NVSS}
images and a literature search reveal that three are ``classical
double'' radio sources (4C~+31.30, 4C~+60.24 = 3C~351, and 4C~+61.20);
the fourth radio source has compact radio emission associated with the
core, but no clear lobes. The remainder of the [OIII]-luminous sources
have no radio emission detected in FIRST, NVSS, or the literature, but
all have upper limits on the FIRST luminosity lying below the
luminosity of the three double-lobed sources. The \citet{MRS93}
prediction was that 6.5 of the sources (50\% of 13) should be double
radio sources, but we found only 3. The Poisson probability to obtain
3 events where 6.5 are expected is 6.9\%.  Thus, our investigation of
BQS-like objects does not allow us to rule out the \citet{MRS93}
result.

In order to obtain some clue as to what causes the large fraction of
radio-loud objects at the high-[OIII] luminosity end of the BQS and
our BQS-like sample, we have constructed a comparison sample of SDSS
quasars using a bright $i$-band cut $i<16.34$, which contains a
similar number of objects as the $B<17$ sample just employed.  This
$i$-band cut removes about 50\% of the sources with $L_{\mathrm{[OIII]
5007}} > 10^{35}$\,W that were present in the $B$-band limited sample,
including one of the three double-lobed sources mentioned above
(4C~+61.20).  Of those objects in the $B$-band limited sample, the
bluest (in $B-i$) are missing in the $i$-band limited one.  Thus,
there is some indication that quasars with a bluer continuum are more
likely to be luminous in [OIII]~5007.  If the link between radio and
narrow-line luminosity identified by \citet[][also see
\citealt{Wileta99}]{MRS93} indeed holds for all quasars, the BQS is
biased towards the inclusion of radio-loud objects because it
preferentially includes objects which are luminous in [OIII]~5007.
This would be consistent with the trend for more luminous quasars to
have bluer optical/UV continua \citep[e.g.]{CGBea99}.  However, an
apparently contradicting observation is that the radio-loud objects
among the SDSS color-selected quasar sample (defined as having $R>1$,
where $R$ is given by Equation~\ref{eq:R}) have \emph{redder} $g-i$
colors than those with $R<1$ \citep[see][]{Ricea01}.  A more detailed
investigation of these connections between [OIII] luminosity, radio
morphology and continuum properties will require the analysis of a
larger sample of bright quasars from the SDSS, and ideally complete
radio imaging of an SDSS-defined bright quasar sample.

\subsection{Summary of SDSS-PG comparison}
\label{s:comp.summary}

In summary, we find the following answers to the two questions posed
at the beginning of \S\ref{s:compPGSDSS}:
\begin{enumerate}
\item The distributions in redshift, \ub, $g-i$, radio
  magnitude $t$, core radio power $P$, and radio-optical ratio $R$ of
  the 39 BQS quasars with $i<15$ inside the SDSS DR3 area are
  indistinguishable from those of the 26 SDSS DR3 quasars passing the
  BQS criteria. Thus, we find no evidence that the incompleteness of
  the BQS is biased with respect to radio or optical properties. We
  compare this finding to previous investigations of the BQS
  incompleteness in \S\ref{s:disc.compBQS} below.
\item The BQS selection effects are predominantly caused by
  application of a flux limit in the $B$-band.  All DR3 quasars
  passing the BQS flux limit also pass the color cut $\ub<-0.44$.
  However, the BQS limits introduce a bias against red objects
  relative to the SDSS sample when considering the colors (both \ub\
  and $g-i$) as a function of $i$-band magnitude (or equivalently, the
  BQS would sample the red part of the quasar population only if it
  was extended to fainter magnitudes).  Thus, the application of
  optical flux limits at different wavelengths introduces a bias that
  is comparable to radio flux limits applied at different wavelengths:
  flat-spectrum (blue) sources are preferentially included in
  high-frequency ($B$-band) selected surveys, while the sample misses
  significant numbers of objects with comparable low-frequency
  ($i$-band) luminosities and steeper (redder) spectra.
\end{enumerate}

Thus, the BQS is representative for quasars which are bright in the
$B$-band, but it is clearly \emph{not} representative even of quasars
which are bright in $i$-band.

\section{Discussion}
\label{s:disc}

\subsection{Comparison to other determinations of BQS completeness}
\label{s:disc.compBQS}

\subsubsection{Color and magnitude incompleteness}
\label{s:disc.compBQS.colormag}

\citet{WP85} presented one of the first investigations of the BQS
completeness.  Their main concerns were a possible color
incompleteness biasing against the inclusion of redder objects, and
that some of the PG limiting magnitudes might actually have been much
fainter than stated by \citet{BQS}.  Similarly, \citet{WCBea00} noted
a possible faint-end overcompleteness of the BQS relative to the
stated flux limits.  Our comparison of SDSS and PG photometry in
\S\ref{s:SDSSphot} and our investigation of the PG completeness
relative to an SDSS-selected sample of UV excess sources in
\S\ref{s:simPG} confirm both of these suggestions: the recalibration
of PG photographic magnitudes we found necessary
(Figure~\ref{f:phot.PGstars_Bfit}) agrees with that suggested by
\citet{WP85} and \citet{WCBea00}, with the PG overestimating the flux
of faint objects.  However, we can rule out a \emph{constant} offset
in the $B$ magnitude such as the one reported by \citet{GMlFea92}.  We
also found that the PG limiting $\ub$ color is in fact closer to
$(\ub)_{\mathrm{lim}}=-0.71$ than to the intended
$(\ub)_{\mathrm{lim}}=-0.44$ (Figure~\ref{f:simPG.Lhood}). This leads to
an increased bias against quasars in the redshift range $0.5<z<1$,
where their median $\ub$ colors make an excursion into the red
(Figure~\ref{f:SDSS_colz}) due to the presence of the MgII emission
line in the $B$ filter \citep[compare][]{WP85}.

Since the $\ub$ distribution of bright quasars peaks at $\ub \approx
-0.7$ (Figure~\ref{f:SDSS_CMDs}), applying the (revised) BQS color cut
$(\ub)_{\mathrm{lim}}<-0.71$ removes about 50\% of the quasars that
should be in the sample.  As long as the color distribution is roughly
symmetric about the mode, this statement is true whatever the
photometric error may be; since the 2-$\sigma$ PG photometric error
actually encompasses the entire width of the $\ub$ distribution at
bright $B$, the detection probability in the BQS will be approximately
independent of $\ub$.  Hence, the incompleteness of the BQS will be
mostly random with respect to $\ub$ color, except for a bias against
quasars with $0.5\la z \la 1.0$, where the median $\ub$ color reaches
$-0.5$.

In this context, it is important to recall that the majority of bright
quasars are found at $z<0.5$ in any case.  This is true for the BQS,
the HEQS \citep{WCBea00} and the SDSS quasar survey. Thus, while the
color-redshift relation of quasars certainly introduces a bias against
intermediate-redshift objects in the BQS, these objects are inherently
rare in the BQS and fainter surveys are necessary for an accurate
determination of the quasar space density at $z>0.5$.

\subsubsection{What is the surface density of bright quasars?}
\label{s:disc.compBQS.surfacedensity}

Short of a determination of the surface density of bright quasars from
the SDSS which requires a detailed analysis of the SDSS completeness
to quasars (\citealp{SOE}; Richards et al. 2005, in preparation), our best
estimate of the incompleteness of the BQS is roughly 50\% because the
corrected $\ub$ cut of $-0.71$ approximately bisects the
\ub\ distribution of bright quasars (Figure~\ref{f:SDSS_CMDs}), in the
sense that the BQS misses half of all quasars which are in fact
brighter than its limiting magnitude. However, the incompleteness in
terms of surface density can be different from this 50\% number
because of Eddington bias, i.e., the scattering of objects in and out
of the sample due to photometric errors.  As discussed in
\S\ref{s:comp.opt.bqslike}, the magnitude of the Eddington bias
depends on the \emph{actual} distribution of quasar magnitudes just
fainter than the survey cutoff and the \emph{actual} distribution of
measurement errors, and cannot be determined exactly based on number
counts exclusively from the bright side of the cutoff.  In fact, our
simulations in \S\ref{s:comp.opt.bqslike} indicate that the number of
BQS objects in the area covered by our ``PG parent sample'' is
consistent with the number expected from the distribution of $B$
magnitudes of the SDSS quasars, so that the surface density deficit
introduced by the color error may have been compensated by
``overcompleteness'' due to Eddington bias.

\citet{WCBea00} have presented the most comprehensive discussion to
date of the BQS incompleteness in terms of surface density. They find
1.48 times as many quasars per square degree, i.e., an incompleteness
of 32\%, but cautioned that the true incompleteness may be as large as
50\% if \citetalias{BQS} underestimated the Eddington bias; this is
indeed the case if our $B$ recalibration is correct, since it makes
the faint plate limits even fainter.  A similar number (53\%$\pm$10\%)
was obtained by \citet{MGVea01} in their most recent analysis of the
First Byurakan Survey, although these authors caution that their
survey area was rather small and results should not be taken as
definite.  \citet{KOKea97} find a similar surface density of UV excess
objects in the Edinburgh-Cape survey as in the PG.  Thus, the report
by \citet{GMlFea92} of a surface density of quasars exceeding that of
the BQS by a factor of three seems to be an outlier.  The most likely
explanation again seems to be small-number statistics in combination
with the revised \ub\ cut and the large photometric errors in the
PG.  We also note that the formal significance of the surface density
difference between EQS and BQS shown in Figure~1 of \citet{GMlFea92}
is less than $2\sigma$.

In this context, let us reconsider what is meant by the term
``effective magnitude limit''.  Figure~\ref{f:intro.BQS.Blim} shows
that the value $B_\mathrm{eff}=16.16$ is close to the area-weighted
median limiting magnitude. However, this is not the same as the
magnitude to which a uniform quasar survey yields the same number of
objects as the entire BQS, with its magnitude limit that is varying
from plate to plate.  If number counts of quasars increase with
limiting apparent magnitude $m$ as $10^{\beta m}$, the effective
limiting magnitude of a survey with a set of limiting magnitudes $m_i$
and total area surveyed to that limiting magnitude $A_i$ is
\begin{displaymath}
\frac{1}{\beta} \log\left(\frac{\sum_i{A_i 10^{\beta m_i}}}{\sum_i
  A_i}\right)
\end{displaymath}
Using any value for $\beta$ from 0.77 \citep{WCBea00} to 0.9
\citep{BQS} with the areas and limiting magnitudes of the PG fields
from \citet{PG} results in an effective limiting magnitude of
$B_\mathrm{eff}=16.20$ (rounded to two decimal places), which would
imply an increase in the expected surface density of the BQS of
7.4\%--8.6\%, and hence a corresponding decrease in all completeness
estimates.  But as with the determination of the Eddington bias in
\S\ref{s:comp.opt.bqslike}, this effective limiting magnitude is only
an \emph{average} value and does not necessarily provide an exact
description of the particular survey under consideration.

In summary, most recent determinations of the surface density of
$B$-bright quasars point to a surface density completeness of the BQS
of 50\% or more. This agrees with our expectation of an incompleteness
of roughly 50\% introduced by the revised BQS color cut
$(\ub)_{\mathrm{lim}}=-0.71$, which bisects the color distribution of
bright quasars, although the color incompleteness will have been
compensated in part or fully by overcompleteness due to Eddington
bias.  In general, the magnitude of the Eddington bias is a random
variable; it is determined by the particular realization of the
probability density functions describing the quasar magnitudes and the
photometric errors, so that techniques like Monte-Carlo simulations or
maximum-likelihood parameter estimation are necessary to assess the
significance of differences in the numbers of quasars found by
different surveys.  We found no strong differences between optical and
radio properties of BQS and BQS-like SDSS quasars.  Thus, we find no
evidence for any systematic effects other than those caused by the
color and magnitude cuts and errors.

\subsection{Science impact of BQS selection effects}
\label{s:disc.impact}

Having shown that bright $B$-band selected quasar surveys (and the PG
survey in particular) are biased to the blue and fail to explore the
reddest part of the quasar population that appears in surveys with
fainter limits, it is important to take stock of the scientific impact
of our reliance on such samples.  This is important because follow-up
studies that require high S/N data, long exposures, and/or good time
sampling must necessarily concentrate on the brightest quasars.

The most obvious impact is on that of our understanding of the mean
quasar spectral energy distribution \citep[e.g.][]{SPNea89,EWMea94}.
The PG sample has a marked influence on these mean SEDs.  Further
exploration of how fainter, redder quasars fit into the mean SED
picture is particularly important given that it is a nearly universal
habit to assume that all quasars have exactly the mean SED (even
though \citet{EWMea94} clearly stress that there is a range of SEDs).

The computation of accretion rates ($\propto
L_\mathrm{Bol}/M_\mathrm{BH}$) generally assumes the mean SED in
determining the bolometric correction rather than an SED tailored for
each object.  Similarly it is important that we reconsider the impact
of the PG sample on the determination of black hole masses.  The two
largest samples of objects on which reverberation analysis has been
applied \citep{WPM99,KSMea00} are (necessarily) dominated by bright
objects and the \citet{KSMea00} sample exclusively consists of PG
quasars.  The issue here is whether the observed BELR size-luminosity
relationship ($R\propto L^{\sim0.5}$) holds for redder quasars.  If
the optical/UV colors of quasars are an indication of the temperature
of the accretion disk, it may not be unreasonable to expect some
scatter in this relationship as a function of color.

The PG sample also impacts our understanding of the X-ray properties
of quasars.  For example, the \citet{EWMea94} sample (19/47 PG
quasars) was specifically chosen to have good S/N in both their {\em
  Einstein} and {\em IUE} spectra and thus are unlikely to be
representative of the typical optical to X-ray ratio, $\alpha_{ox}$.
Furthermore, follow-up studies of quasars in the X-ray are heavily
dominated by the sample of low-$z$ PG quasars defined by
\citet{LFEea97}.  In particular it has been recently suggested that
{\em XMM-Newton} spectra of PG quasars reveals a near
universality of the so-called ``soft X-ray excess'' below 2~keV
\citep{PROea04,PJGea05}.  Again it is important to test if this
feature is indeed universal or only characteristic of bright,
blue quasars.

Finally, we emphasize the obvious influence of the PG sample on the
so-called ``eigenvector'' analysis of quasars initiated by
\citet{BG92}.  This study is based on the analysis of the spectral
properties of the 87 PG quasars with $z<0.5$.  \citet{BG92}
demonstrate that a principal component analysis can be used to define
two key eigenvectors that account for the majority of the variance in
their sample.  While it has been difficult to identify the underlying
physical drivers behind these eigenvectors, one suggestion is that
they might be accretion rate and black hole mass \citep{Bor02}.  If
red quasars lie {\em along} the principal component axes as they are
currently defined, then existing eigenvector analysis is robust to
their addition.  If, however, red quasars lie preferentially off-axis,
then their inclusion may cause a rotation of the axes that will have
implications for the association of the dominant eigenvectors with
underlying physical properties.

\section{Summary}
\label{s:sum}

\begin{enumerate}
\item We have presented separate transformations for stars and quasars
between the SDSS survey system and Landolt photometry on the
Johnson-Kron-Cousins system (\S\ref{s:SDSSphot}).
\item A comparison of SDSS photometry and photographic plate-based
magnitudes from the Palomar-Green (PG) survey \citepalias{PG} of hot
stars (which are not expected to be variable) reveals that the PG
magnitudes require recalibration.  PG photographic magnitudes are
precise in the range $14\la B \la 16$, but too faint by up to 0.2 at
$B\la14$ and $B\ga$ 16. PG photoelectric magnitudes of these objects
agree with SDSS magnitudes to within the photometric errors
(\S\ref{s:phot.sdss.accuracy}).
\item The photographic-plate derived $\ub$ colors of the quasars from
the PG survey \citepalias[the objects included in the Palomar-Green
Bright Quasar Survey, BQS;][]{BQS} are offset by about 0.2 magnitudes
towards the red when compared to SDSS-derived $\ub$ colors.  This
difference is difficult to explain by variability, thus it is an
indication the need to recalibrate the PG $\ub$ colors in addition to
the $B$ magnitudes.  The average and RMS difference in $B$-band
magnitude between the recalibrated PG and SDSS measurements is
consistent with the variability expected over a 30-year epoch
difference (\S\ref{s:phot.sdss.colors}).
\item We have simulated the color selection of objects in the PG
survey from their parent sample as observed by the SDSS DR3 by
computing each object's detection probability from the appropriate PG
limiting magnitude, color, and corresponding errors
(\S\ref{s:simPG}). The parent sample was defined as all objects which
are bluer than the main sequence ($\ub<-0.3$) and covers roughly 3,300
square degrees out of the PG's total survey area of 10,668 square
degrees.  The distributions of $B$ magnitude, $\ub$ color, and
detection probability of the simulated and actual PG survey can only
be made to agree if we adjust the limiting color to
$(\ub)_{\mathrm{lim}}=-0.71$ (in addition to recalibrating the limiting
magnitudes as above), and the photographic color error to
$\sUB=0.24$ (smaller than the value given by \citetalias{PG},
but as given by \citetalias{BQS}).  The offset between our revised
color cut and the cut $(\ub)_{\mathrm{lim}}=-0.44$ intended for quasars
is similar to the offset in the $\ub$ color determined above.  With
the original color cut and error, we predict (\emph{post facto}) that
17,000 main-sequence objects are scattered into the PG's
photometrically defined sample by photometric errors in color and
magnitude.  With our revised limits, the number is reduced to 500
objects, which compares well with the actual number of 1125 objects
rejected over the full PG area.
\item Comparing the properties of 39 BQS quasars inside the SDSS DR3
  area to BQS-like quasars with $B<16.16$ and $\ub<-0.44$ selected
  from the SDSS quasar survey (26 objects), we find no statistically
  significant differences in the distributions in redshift, \ub,
  $g-i$, radio magnitude $t$, core radio power $P$, or radio-optical
  ratio $R$ (\S\ref{s:compPGSDSS}). Moreover, as the $\ub$
  distribution of $B$-bright SDSS quasars peaks at $\ub\approx -0.7$
  and the 2-$\sigma$ $\ub$ color error is about as large as the width
  of the color distribution, inclusion or exclusion in the BQS will be
  essentially random with respect to $\ub$ color, except for a bias
  against objects in the redshift interval $0.5\la z \la 1.0$, where
  the median quasar $\ub$ is much redder than at other redshifts as
  the MgII line passes through the $B$-band.  However, any bright
  quasar survey is limited to low-redshift objects, so that few
  objects from that redshift interval would be included even in the
  absence of color biases.  Thus, we cannot identify any serious
  systematics to the BQS incompleteness.
\item No SDSS quasar brighter than $B=16.16$ is \emph{redder} than
  $\ub=-0.44$, so that the application of the UV excess criterion
  \emph{with the high photometric accuracy of CCD photometry} does not
  remove any objects from the sample that have not already been
  excluded by the application of the $B$-band flux limit.  In fact,
  the number of $B$-bright quasars is so small that the BQS only finds
  objects near the mode of the color distribution and it is necessary to
  reach much fainter $B$ magnitudes than the BQS to find quasars that
  are either much redder or much bluer than the BQS objects.  However,
  the SDSS quasar survey does include objects which are redder in
  $\ub$ than the BQS quasars, but have $i$-band magnitudes similar to
  those of the BQS objects. Thus, while the BQS incompleteness with
  respect to its own selection criteria appears largely random, use of
  the BQS criteria does lead to selection effects compared to a sample
  defined by an $i$-band flux limit, such as the SDSS.  These
  selection effects are predominantly caused by application of a flux
  limit in the $B$-band and not by the application of a UV color
  excess criterion. The $\ub$ color distribution of stars is much
  broader than that of quasars, so that the PG survey \emph{is}
  progressively more biased against inclusion of redder stars.
\item Our recalibration of PG $B$-band magnitudes and colors rules out
  a constant magnitude offset in the BQS magnitudes as reported by
  \citet{GMlFea92}, but agrees with the color and magnitude biases
  perceived by \citet{WP85} and \citet{WCBea00}
  (\S\ref{s:disc.compBQS}).
\item Determination of the surface density of bright quasars from SDSS
  number counts requires detailed analysis of the SDSS quasar
  selection function which is beyond the scope of this paper. Given
  the coincidence of the revised color cut
  $(\ub)_{\mathrm{lim}}=-0.71$ with the mode of the quasar $\ub$
  distribution down to at least $B = 19$, our best estimate of the BQS
  incompleteness is that the BQS misses roughly 50\% of bright quasars
  due to color incompleteness; these losses are at least partially
  canceled by overcompleteness due to Eddington bias.
\item \citet{MRS93} reported that 50\% of the BQS objects at $z<0.5$
  with similarly high $L_{\mathrm{[OIII] 5007}}$ are steep-spectrum
  radio-loud objects.  Of the SDSS quasars at $z<0.5$ and $B<17$,
  thirteen have an [OIII]~5007 narrow-line luminosity of above
  $5\times10^{36}\,$W; three of these are ``classical double'' radio
  sources (\S\ref{s:comp.radio.bias}).  The Poisson probability for
  obtaining 3 sources where 6.5 are expected is 6.9\%.  It is
  necessary to investigate the radio and emission-line properties of a
  larger sample of SDSS-selected quasars in order to obtain a more
  significant rejection or confirmation of the \citet{MRS93} result.
\item Results that have been obtained for PG quasars are commonly
  assumed to be representative for the entire quasar population in
  fundamental issues such as SED shapes, the determination of black
  hole masses via reverberation mapping and related scaling relations,
  quasar X-ray properties, and the physical drivers behind the
  \citet{BG92} eigenvectors.  Given that our work here shows that the
  BQS quasars are just the $B$-bright tip of the quasar iceberg, it
  appears prudent to reassess the impact of the BQS sample selection
  on these results, and to cross-check them for a wider range of
  spectral shapes than encountered among the BQS quasars.
\end{enumerate}

\acknowledgments

We thank Huan Lin for useful conversations, Ari Laor for helpful
comments, and the anonymous referee for constructive criticism.  SJ is
grateful for the hospitality of the Princeton University Observatory,
where this paper was completed. SJ and CS were supported by the US
Department of Energy under contract No.\ DE-AC02-76CH03000.  DPS and
DEVB were partially supported by National Science Foundation grant
AST03-07582, and MAS was supported by grant AST03-07409.  This
research has made use of NASA's Astrophysics Data System as well as
STSDAS/synphot and PyRAF, products of the Space Telescope Science
Institute, which is operated by AURA for NASA.

Funding for the creation and distribution of the SDSS Archive has been
provided by the Alfred P. Sloan Foundation, the Participating
Institutions, the National Aeronautics and Space Administration, the
National Science Foundation, the U.S. Department of Energy, the
Japanese Monbukagakusho, and the Max Planck Society. The SDSS Web site
is http://www.sdss.org/.  The SDSS is managed by the Astrophysical
Research Consortium (ARC) for the Participating Institutions. The
Participating Institutions are The University of Chicago, Fermilab,
the Institute for Advanced Study, the Japan Participation Group, The
Johns Hopkins University, the Korean Scientist Group, Los Alamos
National Laboratory, the Max-Planck-Institute for Astronomy (MPIA),
the Max-Planck-Institute for Astrophysics (MPA), New Mexico State
University, University of Pittsburgh, University of Portsmouth,
Princeton University, the United States Naval Observatory, and the
University of Washington.



\clearpage
\LongTables
\begin{landscape}
\begin{deluxetable}{lrrcccccccccrrrrrr}
\tablecaption{\label{t:BQSinSDSS}Properties of BQS objects in the SDSS DR3 area}
\tabletypesize{\tiny}
\tablewidth{0pt}
\tablehead{\colhead{PG Name} & \colhead{RA\tablenotemark{a}} & \colhead{DEC}
& \colhead{$u$} & \colhead{$g$} & \colhead{$r$} & \colhead{$i$}
& \colhead{$B$} & \colhead{$B$} & \colhead{$\ub$} & \colhead{$\ub$} 
& \colhead{Phot} & \colhead{Targ}
& \colhead{Plt} & \colhead{MJD} & \colhead{Fib} & \colhead{$z$} & \colhead{$z$} \\
\colhead{} & \colhead{(J2000)} & \colhead{(J2000)} & \colhead{} & \colhead{} & \colhead{} & \colhead{} &\colhead{(PG)} & \colhead{(SDSS)} & \colhead{(PG)} &\colhead{(SDSS)} 
& \colhead{(1)} & \colhead{(2)} & \colhead{} & \colhead{} & \colhead{} & \colhead{(SDSS)} & \colhead{(BQS)} 
}
\startdata
0003+15 & 00 05 59.24 & 16 09 49.01 & 15.66 & 15.38 & 15.48 & 15.36 & 15.96 & 15.54 & $-$0.83 & $-$0.61 & 1000 & 0011& \nodata & \nodata & \nodata & \nodata &0.450 \\
0157+00 & 01 59 50.25 & 00 23 40.84 & 15.91 & 15.89 & 15.97 & 15.75 & 15.20 & 16.00 & $-$0.26 & $-$0.79 & 0000 & 0011& 403 & 51871 & 550 & 0.163 &0.164 \\
0844+34 & 08 47 42.47 & 34 45 04.40 & 14.80 & 14.65 & 14.57 & 14.18 & 14.00 & 14.79 & $-$0.75 & $-$0.70 & 1000 & 1001& \nodata & \nodata & \nodata & \nodata &0.064 \\
0921+52 & 09 25 12.85 & 52 17 10.50 & 16.72 & 16.35 & 16.19 & 15.97 & 15.62 & 16.53 & $-$0.35 & $-$0.54 & 0000 & 0011& 767 & 52252 & 418 & 0.035 &0.035 \\
0934+01 & 09 37 01.05 & 01 05 43.71 & 16.72 & 16.61 & 16.54 & 16.31 & 16.29 & 16.74 & \nodata& $-$0.73 & 0010 & 0001& 476 & 52314 & 523 & 0.051 &0.050 \\
0947+39 & 09 50 48.39 & 39 26 50.52 & 16.26 & 16.30 & 16.48 & 15.99 & 16.40 & 16.40 & $-$0.50 & $-$0.84 & 1000 & 0011& 1277 & 52765 & 332 & 0.206 &0.206 \\
0953+41 & 09 56 52.39 & 41 15 22.25 & 14.99 & 14.93 & 14.95 & 14.79 & 15.05 & 15.05 & $-$0.98 & $-$0.76 & 1000 & 1001& \nodata & \nodata & \nodata & \nodata &0.239 \\
1001+05 & 10 04 20.14 & 05 13 00.46 & 16.60 & 16.43 & 16.40 & 16.09 & 16.13 & 16.57 & $-$0.52 & $-$0.69 & 1000 & 0011& 995 & 52731 & 4 & 0.160 &0.161 \\
1012+00 & 10 14 54.90 & 00 33 37.41 & 16.36 & 16.27 & 16.11 & 15.86 & 15.89 & 16.40 & $-$0.44 & $-$0.75 & 1000 & 0111& \nodata & \nodata & \nodata & \nodata &0.185 \\
1022+51 & 10 25 31.28 & 51 40 34.88 & 16.27 & 16.11 & 16.04 & 15.89 & 16.12 & 16.25 & $-$0.37 & $-$0.69 & 0000 & 0011& 1008 & 52707 & 558 & 0.045 &0.045 \\
1049$-$00 & 10 51 51.44 & $-$00 51 17.66 & 16.00 & 15.80 & 15.70 & 15.82 & 15.95 & 15.95 & $-$0.49 & $-$0.66 & 1000 & 0011& 276 & 51909 & 251 & 0.359 &0.357 \\
1103$-$00 & 11 06 31.77 & $-$00 52 52.37 & 16.41 & 16.27 & 16.45 & 16.43 & 16.02 & 16.41 & $-$0.40 & $-$0.71 & 1000 & 0111& \nodata & \nodata & \nodata & \nodata &0.425 \\
1114+44 & 11 17 06.40 & 44 13 33.31 & 16.01 & 15.89 & 15.79 & 15.28 & 16.05 & 16.02 & $-$0.55 & $-$0.72 & 1000 & 0011& 1365 & 53062 & 378 & 0.144 &0.144 \\
1115+08 & 11 18 16.95 & 07 45 58.19 & 16.41 & 16.24 & 16.28 & 16.04 & 15.84 & 16.38 & $-$0.52 & $-$0.69 & 1000 & 0011& 1617 & 53112 & 467 & 1.734 &1.722 \\
1115+40 & 11 18 30.29 & 40 25 54.01 & 15.83 & 15.91 & 16.00 & 15.63 & 16.02 & 16.01 & $-$0.64 & $-$0.87 & 1000 & 0111& 1440 & 53084 & 204 & 0.155 &0.154 \\
1119+12 & 11 21 47.12 & 11 44 18.99 & 15.27 & 15.12 & 14.99 & 14.80 & 14.65 & 15.26 & $-$0.57 & $-$0.70 & 0000 & 1001& \nodata & \nodata & \nodata & \nodata &0.049 \\
1121+42 & 11 24 39.18 & 42 01 45.02 & 16.14 & 15.99 & 15.94 & 15.66 & 16.02 & 16.12 & $-$0.67 & $-$0.70 & 1000 & 0011& 1443 & 53055 & 358 & 0.225 &0.234 \\
1138+04 & 11 41 16.53 & 03 46 59.57 & 17.15 & 17.09 & 17.12 & 17.02 & 16.05 & 17.21 & \nodata& $-$0.77 & 1000 & 0011& 838 & 52378 & 47 & 1.877 &1.876 \\
1148+54 & 11 51 20.46 & 54 37 33.08 & 15.84 & 15.72 & 15.57 & 15.63 & 15.82 & 15.85 & $-$0.50 & $-$0.72 & 1000 & 0111& 1017 & 52706 & 187 & 0.975 &0.969 \\
1151+11 & 11 53 49.27 & 11 28 30.44 & 16.32 & 16.39 & 16.41 & 15.93 & 15.51 & 16.49 & $-$0.57 & $-$0.87 & 1000 & 0011& 1610 & 53144 & 249 & 0.176 &0.176 \\
1206+45 & 12 08 58.01 & 45 40 35.47 & 15.67 & 15.47 & 15.24 & 15.27 & 15.79 & 15.61 & $-$0.66 & $-$0.66 & 1000 & 0011& 1370 & 53090 & 360 & 1.164 &1.158 \\
1216+06 & 12 19 20.93 & 06 38 38.52 & 15.41 & 15.35 & 15.24 & 15.34 & 15.68 & 15.47 & $-$0.42 & $-$0.76 & 1110 & 0001& \nodata & \nodata & \nodata & \nodata &0.334 \\
1226+02 & 12 29 06.70 & 02 03 08.59 & 12.72 & 12.80 & 12.73 & 12.47 & 12.86 & 12.90 & $-$1.18 & $-$0.87 & 1110 & 0001& \nodata & \nodata & \nodata & \nodata &0.158 \\
1244+02 & 12 46 35.25 & 02 22 08.79 & 16.41 & 16.30 & 16.31 & 16.16 & 16.15 & 16.43 & $-$0.49 & $-$0.73 & 0000 & 0011& 522 & 52024 & 173 & 0.048 &0.048 \\
1254+04 & 12 56 59.93 & 04 27 34.39 & 16.47 & 16.24 & 16.06 & 16.07 & 15.84 & 16.39 & $-$0.40 & $-$0.64 & 1000 & 0011& 848 & 52669 & 154 & 1.025 &1.024 \\
1259+59 & 13 01 12.93 & 59 02 06.75 & 15.71 & 15.57 & 15.64 & 15.58 & 15.60 & 15.70 & $-$0.70 & $-$0.70 & 1000 & 0011& 957 & 52398 & 20 & 0.477 &0.472 \\
1307+08 & 13 09 47.00 & 08 19 48.24 & 15.56 & 15.64 & 15.62 & 15.14 & 15.28 & 15.74 & $-$0.82 & $-$0.87 & 1000 & 0011& \nodata & \nodata & \nodata & \nodata &0.155 \\
1322+65 & 13 23 49.52 & 65 41 48.17 & 16.18 & 16.16 & 16.12 & 15.77 & 15.86 & 16.27 & \nodata& $-$0.79 & 1000 & 0011& \nodata & \nodata & \nodata & \nodata &0.168 \\
1329+41 & 13 31 41.13 & 41 01 58.70 & 16.99 & 17.06 & 17.11 & 16.85 & 16.30 & 17.16 & $-$0.38 & $-$0.86 & 1000 & 0011& 1464 & 53091 & 433 & 1.938 &1.930 \\
1338+41 & 13 41 00.78 & 41 23 14.08 & 16.69 & 16.51 & 16.35 & 16.31 & 16.08 & 16.65 & \nodata& $-$0.67 & 1000 & 0011& 1377 & 53050 & 338 & 1.215 &1.219 \\
1351+64 & 13 53 15.83 & 63 45 45.65 & 14.66 & 14.54 & 14.54 & 14.24 & 15.42 & 14.67 & $-$0.53 & $-$0.72 & 1000 & 1001& \nodata & \nodata & \nodata & \nodata &0.087 \\
1352+01 & 13 54 58.68 & 00 52 10.89 & 16.23 & 16.07 & 15.92 & 15.94 & 16.03 & 16.21 & $-$0.47 & $-$0.69 & 1000 & 0011& \nodata & \nodata & \nodata & \nodata &1.117 \\
1411+44 & 14 13 48.33 & 44 00 13.96 & 14.53 & 14.47 & 14.55 & 13.91 & 14.99 & 14.59 & $-$0.81 & $-$0.76 & 1110 & 0001& \nodata & \nodata & \nodata & \nodata &0.089 \\
1415+45 & 14 17 00.82 & 44 56 06.36 & 16.67 & 16.53 & 16.32 & 15.81 & 15.74 & 16.66 & $-$0.79 & $-$0.70 & 0001 & 0001& 1287 & 52728 & 296 & 0.114 &0.114 \\
1426+01 & 14 29 06.57 & 01 17 06.09 & 14.36 & 14.45 & 14.61 & 14.05 & 15.05 & 14.54 & $-$1.32 & $-$0.88 & 1000 & 1001& \nodata & \nodata & \nodata & \nodata &0.086 \\
1427+48 & 14 29 43.07 & 47 47 26.20 & 16.95 & 17.02 & 16.92 & 16.54 & 16.33 & 17.12 & $-$0.43 & $-$0.86 & 1000 & 0011& \nodata & \nodata & \nodata & \nodata &0.221 \\
1440+35 & 14 42 07.47 & 35 26 22.97 & 15.10 & 15.00 & 14.97 & 14.54 & 15.00 & 15.13 & $-$0.53 & $-$0.73 & 1000 & 1001& \nodata & \nodata & \nodata & \nodata &0.077 \\
1444+40 & 14 46 45.94 & 40 35 05.76 & 15.85 & 15.88 & 15.85 & 16.07 & 15.95 & 15.99 & $-$0.58 & $-$0.83 & 1000 & 0011& 1397 & 53119 & 190 & 0.268 &0.267 \\
1501+10 & 15 04 01.20 & 10 26 15.76 & 14.27 & 14.24 & 14.22 & 14.09 & 15.09 & 14.36 & $-$1.16 & $-$0.79 & 1000 & 1001& \nodata & \nodata & \nodata & \nodata &0.036 \\
1512+37 & 15 14 43.07 & 36 50 50.41 & 16.75 & 16.67 & 16.75 & 16.83 & 15.97 & 16.79 & $-$0.46 & $-$0.74 & 1000 & 0111& 1353 & 53083 & 580 & 0.371 &0.371 \\
1522+10 & 15 24 24.53 & 09 58 29.12 & 15.97 & 15.93 & 15.74 & 15.70 & 15.74 & 16.04 & $-$0.50 & $-$0.78 & 1000 & 0011& \nodata & \nodata & \nodata & \nodata &1.321 \\
1534+58 & 15 35 52.41 & 57 54 09.55 & 15.34 & 15.29 & 15.12 & 15.04 & 15.54 & 15.41 & $-$0.54 & $-$0.77 & 0000 & 0011& 615 & 52347 & 108 & 0.030 &0.030 \\
1535+54 & 15 36 38.38 & 54 33 33.31 & 15.58 & 15.20 & 15.01 & 14.77 & 15.31 & 15.37 & $-$0.51 & $-$0.53 & 1000 & 1001& \nodata & \nodata & \nodata & \nodata &0.038 \\
1538+47 & 15 39 34.80 & 47 35 31.31 & 16.55 & 16.08 & 16.05 & 16.14 & 16.01 & 16.27 & \nodata& $-$0.46 & 1010 & 0001& \nodata & \nodata & \nodata & \nodata &0.770 \\
1543+48 & 15 45 30.24 & 48 46 09.07 & 16.56 & 16.44 & 16.44 & 16.46 & 16.05 & 16.57 & $-$0.54 & $-$0.72 & 1000 & 0111& 812 & 52352 & 355 & 0.400 &0.400 \\
1552+08 & 15 54 44.58 & 08 22 21.45 & 16.19 & 15.94 & 15.80 & 15.42 & 16.02 & 16.09 & $-$0.45 & $-$0.62 & 1000 & 0011& \nodata & \nodata & \nodata & \nodata &0.119 \\
1612+26 & 16 14 13.20 & 26 04 16.20 & 16.38 & 16.38 & 16.17 & 15.61 & 16.00 & 16.49 & $-$0.69 & $-$0.81 & 0000 & 0011& 1393 & 52824 & 35 & 0.131 &0.131 \\
1630+37 & 16 32 01.12 & 37 37 50.01 & 16.13 & 16.03 & 15.93 & 15.80 & 15.96 & 16.15 & $-$0.28 & $-$0.73 & 1000 & 0011& 1173 & 52790 & 198 & 1.476 &1.471 \\
1704+60 & 17 04 41.38 & 60 44 30.50 & 15.63 & 15.43 & 15.24 & 15.18 & 15.90 & 15.58 & $-$0.42 & $-$0.66 & 1000 & 0011& 353 & 51703 & 377 & 0.372 &0.371 \\
2130+09 & 21 32 27.82 & 10 08 19.16 & 14.73 & 14.70 & 14.80 & 14.28 & 14.62 & 14.81 & \nodata& $-$0.79 & 1000 & 1001& \nodata & \nodata & \nodata & \nodata &0.061 \\
2214+13 & 22 17 12.26 & 14 14 20.89 & 14.55 & 14.51 & 14.50 & 13.99 & 14.98 & 14.62 & \nodata& $-$0.78 & 1000 & 1001& \nodata & \nodata & \nodata & \nodata &0.067 \\
2233+13 & 22 36 07.68 & 13 43 55.32 & 16.30 & 16.28 & 16.28 & 16.41 & 16.04 & 16.39 & $-$0.62 & $-$0.80 & 1000 & 0011& 739 & 52520 & 388 & 0.326 &0.325 \\
\enddata
\tablenotetext{a}{Unless indicated otherwise, all quantities are taken
from the SDSS photometric and spectroscopic catalogs.}
\tablecomments{We use PG object names, $B$ (PG) and $z$ (BQS) for the
BQS objects \citep{BQS} as listed in the on-line tables from
\citet[][CDS VizieR catalog J/AJ/108/1163]{KSSea94}, which includes
the corrections to the BQS catalog given by \citet{PG}. The previously
unpublished photographic $\ub$ colors from \citet{BQS} are given in
column $\ub$ (PG) when available.  $B$ (SDSS) and $\ub$ (SDSS) have
been obtained from SDSS $u$ and $g$ using the transformations given in
\S\ref{s:phot.qsotrans}.  The columns \emph{Phot} and \emph{Targ} are
sets of flags which give additional information about SDSS photometry
and target selection in the {\tt BEST} version of the photometric
catalogs. (1) \emph{Phot} contains information (in this order) on
whether the source was a point source, saturated in any of the SDSS
bands, had a fatal cosmetic error in quasar target selection, and
whether it had a nonfatal cosmetic error in quasar target
selection. (2) \emph{Targ} contains information on whether the object
was flagged as quasar candidate by color selection but is too bright
to be observed with the spectrograph, targeted as FIRST source,
targeted by color selection, and finally a ``completeness'' flag. This
``completeness'' flag is the logical OR of the flags indicating a
fatal or non-fatal cosmetic error, of those identifying the object as
color- or FIRST-selected target, and of the flag indicating a
color-selected target that is too bright for inclusion in the
spectroscopic survey. The ``completeness'' flag would be 0 if an
object had neither been recognized as quasar target nor rejected only
due to cosmetic problems, causing the object to be missed by SDSS
target selection; it is 1 for all objects in the table, indicating
that every BQS quasars was either targeted, or we understand why it
was not targeted. The columns Plt, MJD, Fib give the plate/MJD/fiber
combination identifying the SDSS spectrum, where available. All
spectra of BQS objects were classified as quasar by the automated
pipeline, with the redshift given as $z$ (SDSS).}
\end{deluxetable}
\clearpage
\end{landscape}

\end{document}